\documentclass[%
 aip,
 pof,
 amsmath,amssymb,
 preprint,
 numerical
]{revtex4-1}

\usepackage{amsmath}
\usepackage{epsfig}
\usepackage{graphicx}
\usepackage{dcolumn}
\usepackage{bm}

\newcommand{\ud}{\mathrm{d}}
\newcommand{\ue}{\mathrm{e}}
\newcommand{\uF}{\mathrm{F}}
\newcommand{\ug}{\mathrm{g}}
\newcommand{\ui}{\mathrm{i}}
\newcommand{\umax}{\mathrm{max}}

\begin{document}

\title{A theoretical and numerical study of gravity driven coating flow on cylinder and sphere: fingering instability}

\author{Shuo Hou}
 \email{hs180934@163.com.}

\affiliation{
School of Aeronautic Science and Engineering, Beihang University, Beijing, China
}

\begin{abstract}
To find the regularities of formed fingers in gravity driven coating flows on upper cylinder and sphere, the mathematical formulation to model the fingering instability on cylindrical or spherical surface which consists of a capillary wave equation and a linear perturbation equation is constructed, based on the leading order governing equation in standard cylindrical or spherical coordinate system. A disjoining pressure model is introduced to simulate the partial wetting process near moving contact line. The fingering number in high $Bo$ coating flow is focused on from a linear perspective. Using an asymptotic theory, the high $Bo$ limits of the linear perturbation equations for both the cylindrical and spherical problems are proved to degenerate into a common eigenvalue problem. Two analytical formulae concerning the most unstable wave number which can be described as two power laws are derived via a method of modal analysis for the cylindrical and spherical problem, respectively. These formulae provide a succinct method to estimate the fingering number if the onset of the fingering instability is known. A method of transient growth analysis which makes no high $Bo$ assumptions is used to verify the accuracy of the asymptotic theory and modal analysis. In particular, an asymptotic behavior of the most unstable wave number as $Bo$ increases is highlighted. We show that the linear growth rate may be improved by the disjoining pressure, which can be attributed to the effect of partial wetting, but the most unstable wave number is insensitive to the wetting model.
\end{abstract}

\maketitle

\section{\label{sec:intro}Introduction}

The fingering of coating flows down a slope is an example that is familiar in daily life. However, the comprehensive experiments and scientific explanations concerning this phenomenon were presented about three decades ago, as first introduced by \citet{Huppert1982}, then by \citet{Silvi1985} and \citet{Bruyn1992}. \citet{Troian1989} first described the theoretical mechanism for the fingering instability using an appropriate linear analysis, then this mechanism is further validated and extended by \citet{Spaid1996}. They pointed out that the fingering phenomena can be described as transverse instability of a moving capillary wave. The mechanism for the instability involves a subtle interplay between the fluid far from the moving front, where surface tension is irrelevant, and the fluid near the contact line, where surface tension dominates. A capillary ridge is always formed in the surface-tension dominated region, which is responsible for the subsequent instability. Since the initial experiments and analysis on viscous flows down an inclined plane, studies have focused on more features of the fingering instability as well as other experimental configurations, such as spin coating \citep{Melo1989}, Marangoni forcing \citep{Kataoka1997} and surfactant driven spreading \citep{Warner2004}. It is believed that these phenomena belong to a same class of problems with different driven forces. The common feature is that the formation of a capillary ridge near the contact line, which is unstable to spanwise disturbances.

There is another direction to extend the driven coating problems on a slope. The shape of solid substrate can be generalized from a plane to a curved surface. Apparently the free-surface flow on curved surface may have more abundant characteristics than planar problem, because planar surface belongs to a special curved surface. \citet{Schwartz1995} and later \citet{Kalliadasis2000} showed some features of steady film profile on a two-dimensional curved section. Compared to the planar problem, the capillary wave may also exist on curved surface, and the fingering instability may occur if the moving front of this capillary wave is disturbed. It has been reported that the fingers exist on the outside of a horizontal \citep{Evans2005} or vertical \citep{Smolka2011} cylinder by using experimental and numerical methods. Recently \citet{Takagi2010} presented the flow and stability of viscous films down the outside of a cylinder and a sphere via some well-designed experiments. They observed the fingering phenomena on both the cylinder and sphere, but only the instability data on spherical surface is given in detail. They pointed out that there is regularity concerning the wavelength of the formed fingers, but they did not present a quantitative relationship for this wavelength which describes the distance between two fingers.

It should be noted that there are three important parameters to describe the fingering instability -- fingering number, fingering pattern and the growth rate of the fingers' length. The linear theory proposed by \citet{Troian1989} can give a finite-wavelength most unstable mode and corresponding growth rate, which can be used to model fingering number and growth, from a linear perspective. But the fingering pattern can not be solved directly using a linear theory. Fingering instability is essentially a kind of nonlinear instability of an unsteady spreading capillary wave, and the linear theory can only model the early stage of the instability process. The growth rate at later stage and specially the fingering pattern are determined by the nonlinear evolution. Moreover, if the solid surface is sufficiently dry, the wetting parameters such as contact angle may play a role on the free-surface flow near the moving contact line. Consequently the fingering pattern and growth rate are also determined by the surface physicochemical properties, which introduce more complications. The mechanism of fingering instability in nonlinear stage is not yet clear at present, even for the canonical problem concerning coating flow on a slope. However, the weakly nonlinear analysis, direct numerical simulation and many experiments showed that the fingering number will not be changed observably after the fingers are formed, due to the nonlinear suppression in the later stage \citep{Craster2009}. It is believed that the most unstable Fourier mode in linear theory can give reliable reference for fingering number in a practical coating flow problem.

The linear analysis in \citet{Troian1989}'s original paper belongs to one kind of modal theory. There are two main assumptions in classic modal analysis. First, the base state of the linear instability -- the moving capillary wave is quasi-steady, which means that the capillary wave is steady in a reference frame traveling at a uniform wave speed. Second, the linearized operator is normal and the eigenvectors are orthogonal, so that the linear perturbation equation can degenerate into an eigenvalue problem. But both of the two assumptions may not be satisfied strictly in a practical case. For the canonical problem of coating flow on a slope, according to Huppert's similarity solution \citep{Huppert1982} in outer region, the front thickness varies with time slowly when the effect of initial condition can be neglected, thus the moving capillary wave is essentially unsteady. Moreover, the presence of a deformable free-surface gives rise to capillary wave that are spatially nonuniform. This spatial inhomogeneity produces weakly nonnormal disturbance operators, which is an initial value problem related to initial conditions of disturbance. To overcome these imperfections, the non-modal method is developed by solving the linear perturbation equation directly. The initial disturbance can be set to arbitrary type, and the base state can be quasi-steady or completely unsteady. Non-modal method mainly observes the disturbance energy at a moment, also known as transient growth analysis (TGA) \citep{Matar2002,Li2014}. The transient growth rate should be calculated from the evolution of the disturbance energy, which is different from directly using the eigenvalue as a constant growth rate in classic modal analysis.

We are concerned here with the fingering instability of coating flows on the outside of a horizontal cylinder and a sphere. The flow is mainly driven by gravity. Due to the aforementioned difficulties in nonlinear mechanism, we are only concerned with the fingering number in this paper. It can be described as an interesting question: how many fingers may be formed on upper cylinder or sphere if the onset of the fingering instability is a known priori. The quantitative profiles of the two-dimensional capillary waves on cylinder and axisymmetric capillary waves on sphere have been studied in detail in our previously submitted paper\citep{Hou2017}. These capillary waves can be used as the base state for subsequent fingering on corresponding surface. Actually the linear perturbation equations on cylinder and sphere can be derived directly from the complete governing equations, and we will present the forms of these equations at first. In our previous paper, we emphasized the importance of a Bond number and found some asymptotic behaviors under high Bond number condition using an asymptotic theory. In this paper, this asymptotic theory will be extended to analyze the spanwise instability of the two-dimensional and axisymmetric capillary waves for high Bond number flow. The linear perturbation equations will degenerate into a common eigenvalue problem which is similar to the planar coating flow, so that the classic modal analysis can be applied to obtain some theoretical formulae. We will show that, there are exactly regularities in the formed fingers, some power laws concerning the most unstable wave number exist on both the cylinder and sphere, if the asymptotic conditions can be satisfied. These formulae provide a succinct method to estimate the fingering number for a practical coating flow problem, when the beginning of the destabilization is observed. In addition, because of the imperfections in the modal analysis, the TGA method is used to provide linear growth results and compared to the deductions of asymptotic theory and modal analysis.

The organization of this paper is as follows. In \ref{sec:theo}, the leading order governing equations for coating flow on cylindrical and spherical surfaces are presented. The mathematical formulation which consists of a capillary wave equation and a linear perturbation equation can be derived from the complete governing equation to model the fingering instability on cylinder and sphere, respectively. For partial wetting cases, a disjoining pressure model is introduced into the governing equations. In \ref{sec:ast}, an asymptotic theory is used to study the high Bond number limit of the linear perturbation equations and to derive some theoretical results for the fingering number which may be formed. Some solution techniques for numerically solving the eigenvalue equation introduced by the asymptotic theory and the original linear perturbation equations are discussed in \ref{sec:num}. In \ref{sec:resdis}, we show the numerical results of the eigenvalues and eigenfunctions solved from the eigenvalue problem via a method of modal analysis. The linear growth is presented using the approach of transient growth analysis. The comparisons between the modal analysis, transient growth analysis and available experimental results are discussed. We highlight the asymptotic behavior of the most unstable wave number as the Bond number increases. The effect of disjoining pressure on the linear growth is also discussed. Section \ref{sec:con} is the concluding remarks of the present work.

\section{\label{sec:theo}Mathematical formulation}

\subsection{Governing equations and nondimensionalization}

\begin{figure}
  \includegraphics[width=0.49\textwidth]{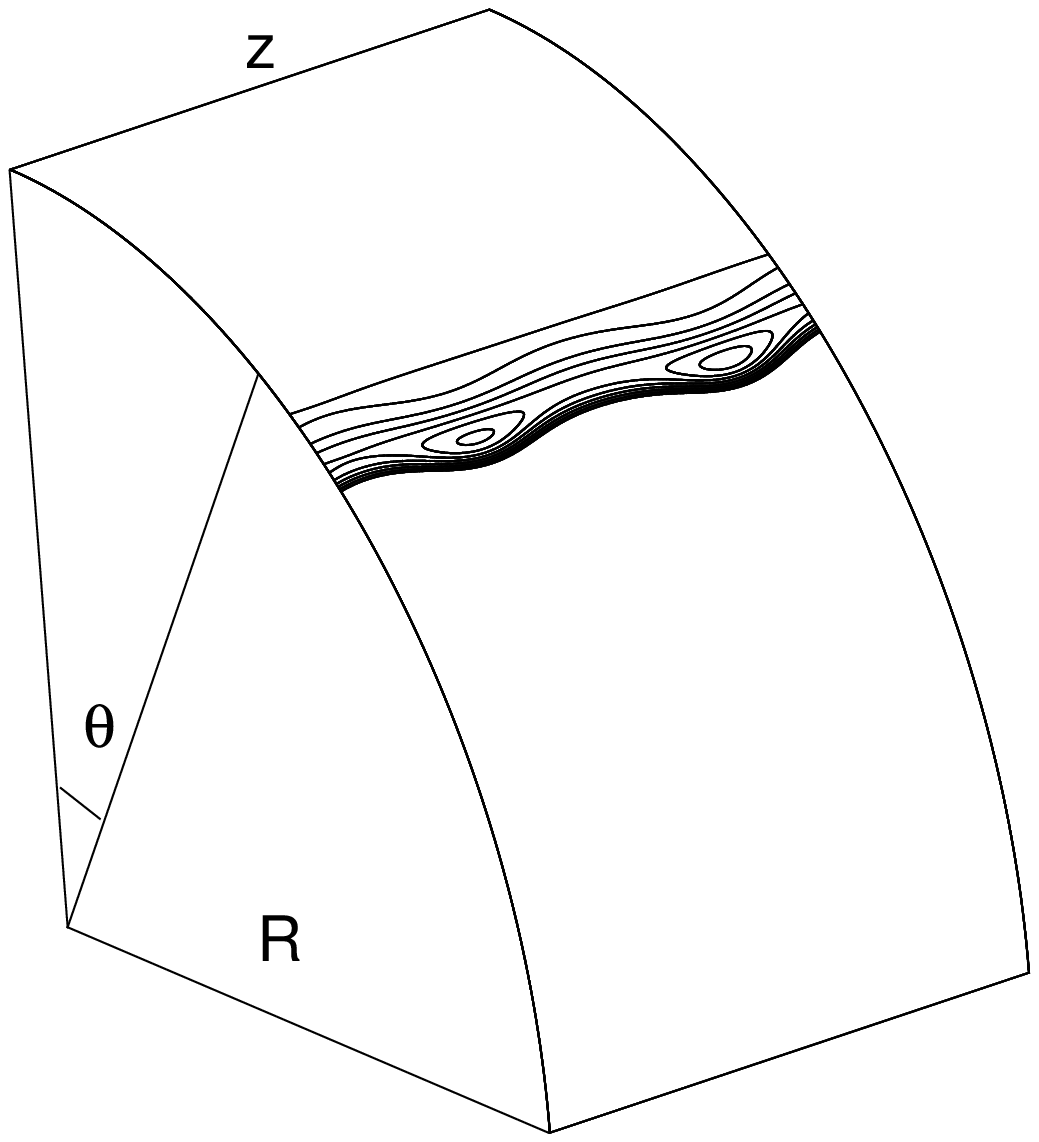}
  \includegraphics[width=0.49\textwidth]{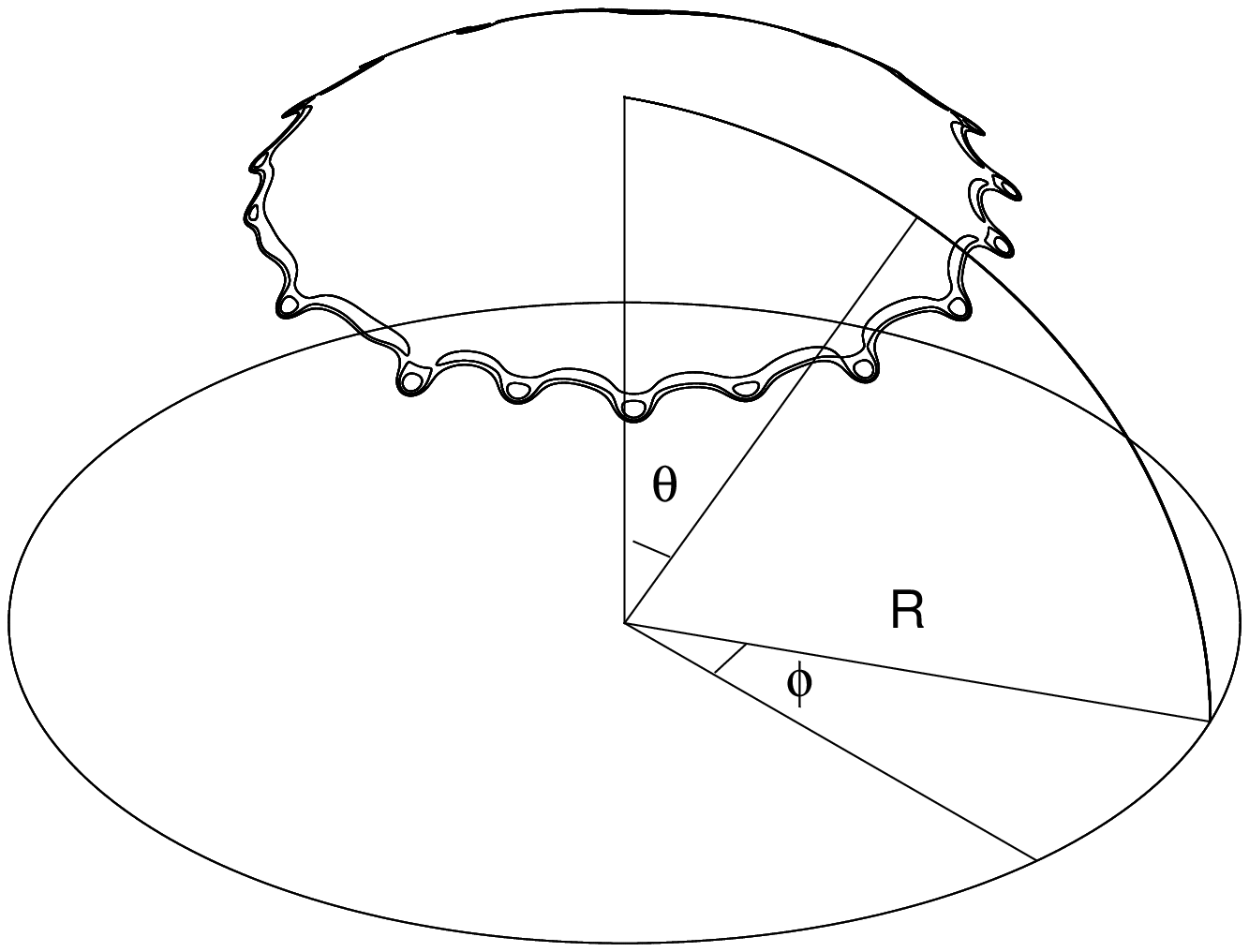}
  \centerline{(a)\hspace{0.45\textwidth}(b)}
  \caption{\label{fig:diagram} The schematic diagram of the fingering instability at early stage on (a) cylinder and (b) sphere.}
\end{figure}

We study the evolution of a Newtonian liquid film on the upper part of a solid cylinder or sphere, as it drips down along the surface. The axis of the cylinder is horizontal. The schematic diagrams concerning the early stage of the fingering instability are shown in Fig.~\ref{fig:diagram}(a) and Fig.~\ref{fig:diagram}(b), for cylindrical and spherical problem, respectively. A regular perturbation method which is similar to the lubrication theory for coating flow on a slope can be used to derive the leading order governing equation on cylindrical or spherical surface \citep{Hou2017}. For cylindrical problem, the governing equation is
\begin{equation}\label{eq:cydge}
\frac{\partial h}{\partial t}+\frac{1}{R}\frac{\partial}{\partial\theta}[\frac{h^{3}}{3\mu}(-\frac{1}{R}\frac{\partial p_{s}}{\partial\theta}+
\rho g\sin\theta)]+\frac{\partial}{\partial z}[\frac{h^{3}}{3\mu}(-\frac{\partial p_{s}}{\partial z})]=0
\end{equation}
and for spherical problem, it is
\begin{equation}\label{eq:spdge}
\frac{\partial h}{\partial t}+\frac{1}{R\sin\theta}\frac{\partial}{\partial \theta}[\sin\theta\frac{h^{3}}{3\mu}(-\frac{1}{R}\frac{\partial p_{s}}{\partial
\theta}+\rho g\sin\theta)]+\frac{1}{R\sin\theta}\frac{\partial}{\partial \phi}[\frac{h^{3}}{3\mu}(-\frac{1}{R\sin\theta}\frac{\partial p_{s}}{\partial \phi})]=0
\end{equation}
where $\rho$ is fluid density, $\mu$ is viscosity, $g$ is gravity acceleration and $R$ is the radius of the cylinder or sphere. $(\theta, z)$ and $(\theta, \phi)$ is the standard cylindrical coordinates and spherical coordinates on the respective surface, $t$ is time and $h$ is the local thickness of liquid film. The Young-Laplace pressure $p_s$ can be expressed as
\begin{equation}\label{eq:cylyp}
p_{s}=-\sigma(\frac{1}{R^{2}}h+\frac{1}{R^{2}}\frac{\partial^{2}h}{\partial\theta^{2}}+\frac{\partial^{2}h}{\partial z^{2}})
\end{equation}
for cylindrical problem, and
\begin{equation}\label{eq:splyp}
p_{s}=-\sigma(\frac{2}{R^{2}}h+\frac{1}{R^{2}\tan\theta}\frac{\partial h}{\partial\theta}+
\frac{1}{R^{2}}\frac{\partial^{2}h}{\partial\theta^{2}}+\frac{1}{R^{2}\sin^{2}\theta}\frac{\partial^{2}h}{\partial\phi^{2}})
\end{equation}
for spherical problem, where $\sigma$ is surface tension of the fluid. The details of the derivations can be found in \citet{Hou2017}.

The following scales are used to nondimensionaliz the governing equations on cylindrical or spherical surface
\begin{subequations}
\begin{equation}\label{eq:scaleh}
h=Hh^{*},t=\frac{3\mu R}{\rho gH^{2}}t^{*},p=\rho gRp^{*}
\end{equation}
\begin{equation}\label{eq:scalez}
z=Rz^{*}
\end{equation}
\end{subequations}
where $H$ is thickness scales of the liquid film. Then, the dimensionless evolution equation for cylindrical problem can be obtained as
\begin{subequations}\label{eq:cyndge}
\begin{equation}
\frac{\partial h^{*}}{\partial t^{*}}+\frac{\partial}{\partial\theta}[h^{*3}(-\frac{\partial p_{s}^{*}}{\partial\theta}+\sin\theta)]+
\frac{\partial}{\partial z^{*}}(-h^{*3}\frac{\partial p_{s}^{*}}{\partial z^{*}})=0
\end{equation}
\begin{equation}\label{eq:cyndp}
p_{s}^{*}=-\frac{1}{Bo}(h^{*}+\frac{\partial^{2}h^{*}}{\partial\theta^{2}}+\frac{\partial^{2}h^{*}}{\partial z^{*2}})
\end{equation}
\end{subequations}
Analogously the dimensionless forms for spherical problem are
\begin{subequations}\label{eq:spndge}
\begin{equation}
\frac{\partial h^{*}}{\partial t^{*}}+\frac{1}{\sin\theta}\frac{\partial}{\partial\theta}[\sin\theta h^{*3}(-\frac{\partial p_{s}^{*}}{\partial \theta}+\sin\theta)]+\frac{1}{\sin\theta}\frac{\partial}{\partial\phi}(-\frac{h^{*3}}{\sin\theta}\frac{\partial p_{s}^{*}}{\partial\phi})=0
\end{equation}
\begin{equation}\label{eq:spndp}
p_{s}^{*}=-\frac{1}{Bo}(2h^{*}+\frac{1}{\tan\theta}\frac{\partial h^{*}}{\partial\theta}+
\frac{\partial^{2}h^{*}}{\partial\theta^{2}}+\frac{1}{\sin^{2}\theta}\frac{\partial^{2}h^{*}}{\partial\phi^{2}})
\end{equation}
\end{subequations}
In the dimensionless expressions of Young-Laplace pressure, the definition of Bond number $Bo$ is identical for both the cylindrical and spherical problems
\begin{equation}\label{eq:bo}
Bo=\frac{\rho gR^{3}}{\sigma H}
\end{equation}

\subsection{Two-dimensional capillary wave and linear perturbation on cylinder}

The two-dimensional form of governing equation (\ref{eq:cyndge}) can be obtained via a condition of $\frac{\partial h}{\partial z}=0$
\begin{equation}\label{eq:tdeq}
\frac{\partial h}{\partial t}+\frac{\partial}{\partial \theta}\{h^{3}[\frac{1}{Bo}\frac{\partial}{\partial \theta}(h+
\frac{\partial^{2}h}{\partial \theta^{2}})+\sin\theta]\}=0
\end{equation}
The dimensionless asterisk signs are neglected hereinafter. The two-dimensional flow can be realized by setting a two-dimensional initial condition for (\ref{eq:cyndge}). For a uniformly distributed initial liquid film which is located at the top of the cylinder and is symmetrical about the vertical plane, the dimensionless initial condition is
\begin{equation}\label{eq:cwic1}
\begin{aligned}
  h_{\ui}(\theta) & =1 \qquad 0\leq\theta\leq\theta_{\ui} \\
  h_{\ui}(\theta) & =b \qquad \theta_{\ui}<\theta\leq\pi
\end{aligned}
\end{equation}
The initial film thickness $H$ is used as the thickness scale, $\theta_\ui$ is a preset angle where the initial front is located, $b$ is dimensionless thickness of a precursor film which is assumed ahead of the initial front to overcome the contact line singularity. Alternatively a continuous hyperbolic tangent function can be used to approximate initial profile (\ref{eq:cwic1})
\begin{equation}\label{eq:cwic2}
h_{\ui}(\theta)=\frac{1+b}{2}-\frac{1-b}{2}\tanh[a(\theta-\theta_{\ui})] \qquad 0\leq\theta\leq\pi
\end{equation}
where $a$ is a coefficient for controlling the width of transition zone. The boundary conditions for (\ref{eq:tdeq}) are
\begin{equation}\label{eq:cwbc}
h_{\theta},h_{\theta\theta\theta}=0\mid_{\theta=0,\pi}
\end{equation}
in which the subscripts represent derivatives.

It has been demonstrated that a two-dimensional capillary wave can be solved directly from (\ref{eq:tdeq}) \citep{Hou2017}. This traveling wave solution is considered as the base state for the subsequent evolution. Consider a perturbation is superimposed onto the moving front of the base state $h_0$
\begin{equation}\label{eq:cypertod}
h(\theta,z,t)=h_{0}(\theta,t)+\epsilon G(\theta,z,t)
\end{equation}
We think of $\epsilon\ll1$ and the perturbation function $G$ is $O(1)$. Substituting (\ref{eq:cypertod}) into (\ref{eq:cyndge}), and linearization of (\ref{eq:cyndge}) with respect to $O(\epsilon)$ yields the following fourth-order linear partial differential equation
\begin{subequations}
\begin{equation}\label{eq:cypert}
\frac{\partial G}{\partial t}+\frac{\partial}{\partial\theta}[-h_{0}^{3}\frac{\partial p_{\ug}}{\partial \theta}+3h_{0}^{2}(-\frac{\partial p_{s}}{\partial \theta}+\sin\theta)G]+\frac{\partial}{\partial z}(-h_{0}^{3}\frac{\partial p_{\ug}}{\partial z})=0
\end{equation}
\begin{equation}\label{eq:cyps}
p_{s}=-\frac{1}{Bo}(h_{0}+\frac{\partial^{2}h_{0}}{\partial\theta^{2}})
\end{equation}
\begin{equation}\label{eq:cypg}
p_{\ug}=-\frac{1}{Bo}(G+\frac{\partial^{2}G}{\partial\theta^{2}}+\frac{\partial^{2}G}{\partial z^{2}})
\end{equation}
\end{subequations}
Since the two-dimensional capillary wave $h_0$, does not depend on the transverse variable $z$, (\ref{eq:cypert}) can be Fourier transformed in $z$
\begin{subequations}\label{eq:cyperta}
\begin{equation}\label{eq:cypertq}
\frac{\partial G}{\partial t}+\frac{\partial}{\partial\theta}[-h_{0}^{3}\frac{\partial p_{\ug}}{\partial \theta}+3h_{0}^{2}(-\frac{\partial p_{s}}{\partial\theta}+\sin\theta)G]+h_{0}^{3}q^{2}p_{\ug}=0
\end{equation}
\begin{equation}\label{eq:cypgq}
p_{\ug}=-\frac{1}{Bo}(G+\frac{\partial^{2}G}{\partial\theta^{2}}-q^{2}G)
\end{equation}
\end{subequations}
in which $q$ is wave number in $z$ direction, $p_s$ is same as (\ref{eq:cyps}). Equation (\ref{eq:cypertq}) is the final linear perturbation equation for a given normal mode $G(\theta,t)e^{iqz}$, with the boundary conditions
\begin{equation}\label{eq:pertbc}
G_{\theta},G_{\theta\theta\theta}=0\mid_{\theta=0,\pi}
\end{equation}

\subsection{Axisymmetric capillary wave and linear perturbation on sphere}

The axisymmetric equation on spherical surface can be derived by using $\frac{\partial h}{\partial\phi}=0$ to simplify the governing equation (\ref{eq:spndge})
\begin{equation}\label{eq:axeq}
\frac{\partial h}{\partial t}+\frac{1}{\sin\theta}\frac{\partial}{\partial \theta}\{\sin\theta h^{3}[\frac{1}{Bo}\frac{\partial}{\partial \theta}(2h+\frac{1}{\tan\theta}\frac{\partial h}{\partial\theta}+\frac{\partial^{2}h}{\partial \theta^{2}})+\sin\theta]\}=0
\end{equation}
The initial conditions and boundary conditions for (\ref{eq:axeq}) are identical to the cylindrical problem. The only difference is that the range of polar angle $\theta$ is natively from 0 to $\pi$, and the additional symmetry plane is not necessary for spherical problem.

An axisymmetric capillary wave can also be solved directly from (\ref{eq:axeq}). With this capillary wave as base state, the fingering instability is examined by superimposing perturbation onto the front
\begin{equation}\label{eq:sppertod}
h(\theta,\phi,t)=h_{0}(\theta,t)+\epsilon G(\theta,\phi,t)
\end{equation}
where $\epsilon\ll1$ and $G$ is perturbation function. Substituting (\ref{eq:sppertod}) into (\ref{eq:spndge}) and keeping only those terms that are $O(\epsilon)$, the equation for the perturbation $G$ is
\begin{subequations}
\begin{equation}\label{eq:sppert}
\frac{\partial G}{\partial t}+\frac{1}{\sin\theta}\frac{\partial}{\partial \theta}\{\sin\theta[-h_{0}^{3}\frac{\partial p_{\ug}}{\partial\theta}+3h_{0}^{2}(-\frac{\partial p_{s}}{\partial\theta}+\sin\theta)G]\}+
\frac{1}{\sin\theta}\frac{\partial}{\partial \phi}(-\frac{h_{0}^{3}}{\sin\theta}\frac{\partial p_{\ug}}{\partial\phi})=0
\end{equation}
\begin{equation}\label{eq:spps}
p_{s}=-\frac{1}{Bo}(2h_{0}+\frac{1}{\tan\theta}\frac{\partial h_{0}}{\partial\theta}+\frac{\partial^{2}h_{0}}{\partial\theta^{2}})
\end{equation}
\begin{equation}\label{eq:sppg}
p_{\ug}=-\frac{1}{Bo}(2G+\frac{1}{\tan\theta}\frac{\partial G}{\partial\theta}+\frac{\partial^{2}G}{\partial \theta^{2}}+\frac{1}{\sin^{2}\theta}\frac{\partial^{2}G}{\partial\phi^{2}})
\end{equation}
\end{subequations}
Due to the implicit periodic condition in azimuthal direction
\begin{equation}\label{eq:pedbc}
h(\theta,\phi+2\pi,t)=h(\theta,\phi,t)
\end{equation}
the perturbation function $G$ can be expanded using Fourier series in azimuthal direction
\begin{equation*}
G(\theta,\phi,t)=\sum_{q=-\infty}^{+\infty}G(\theta,q,t)e^{iq\phi}
\end{equation*}
$q$ is wave number and should be integer. Since (\ref{eq:sppert}) is linear and the capillary wave solution $h_0$ does not depend on the azimuthal angle, we can study the stability of a single Fourier mode and use superposition principle to obtain the evolution of an arbitrary initial perturbation. The equation for the perturbation function $G(\theta,q,t)$ can be derived by substituting $G(\theta,t)e^{iq\phi}$ into (\ref{eq:sppert}) and (\ref{eq:sppg})
\begin{subequations}\label{eq:spperta}
\begin{equation}\label{eq:sppertq}
\frac{\partial G}{\partial t}+\frac{1}{\sin\theta}\frac{\partial}{\partial \theta}\{\sin\theta[-h_{0}^{3}\frac{\partial p_{\ug}}{\partial\theta}+3h_{0}^{2}(-\frac{\partial p_{s}}{\partial\theta}+\sin\theta)G]\}+\frac{h_{0}^{3}q^{2}p_{\ug}}{\sin^{2}\theta}=0
\end{equation}
\begin{equation}\label{eq:sppgq}
p_{\ug}=-\frac{1}{Bo}(2G+\frac{1}{\tan\theta}\frac{\partial G}{\partial\theta}+\frac{\partial^{2}G}{\partial \theta^{2}}-\frac{q^{2}G}{\sin^{2}\theta})
\end{equation}
\end{subequations}
with the identical boundary conditions compared to the cylindrical problem
\begin{equation*}
G_{\theta},G_{\theta\theta\theta}=0\mid_{\theta=0,\pi}
\end{equation*}

\subsection{The wetting model}

For partial wetting process \citep{Gennes1985} which commonly exists in the coating flows, a disjoining pressure model can be introduced to approximate the submicroscopic intermolecular forces near the contact line where a finite contact angle exists \citep{Schwartz1998}
\begin{subequations}
\begin{equation}\label{eq:ddjp}
\Pi=B[(\frac{h_{\mathrm{min}}}{h})^{n}-(\frac{h_{\mathrm{min}}}{h})^{m}]
\end{equation}
\begin{equation}\label{eq:djpco}
B=\frac{(n-1)(m-1)}{2h_{\mathrm{min}}(n-m)}\sigma\theta_{\ue}^{2}
\end{equation}
\end{subequations}
in which the exponents $n$ and $m$ are positive constants with $n>m>1$. The constant $B$ is positive and has the dimension of pressure, $\theta_\ue$ is the equilibrium contact angles and $h_{min}$ is a stable film thickness. This disjoining pressure is considered as a modification to the Young-Laplace pressure at free-surface
\begin{equation}\label{eq:totip}
p_{t}=p_{s}-\Pi
\end{equation}
The expressions of governing equation in partial wetting cases can be derived by replacing $p_s$ by $p_t$.

For two-dimensional or axisymmetric evolution, the dimensionless equations become
\begin{equation}\label{eq:tddjpeq}
\frac{\partial h}{\partial t}+\frac{\partial}{\partial \theta}\{h^{3}[\frac{1}{Bo}\frac{\partial}{\partial \theta}(h+
\frac{\partial^{2}h}{\partial \theta^{2}}+\Pi_{\ud})+\sin\theta]\}=0
\end{equation}
for cylinder, and
\begin{equation}\label{eq:axdjpeq}
\frac{\partial h}{\partial t}+\frac{1}{\sin\theta}\frac{\partial}{\partial \theta}\{\sin\theta h^{3}[\frac{1}{Bo}\frac{\partial}{\partial \theta}(2h+\frac{1}{\tan\theta}\frac{\partial h}{\partial\theta}+\frac{\partial^{2}h}{\partial \theta^{2}}+\Pi_{\ud})+\sin\theta]\}=0
\end{equation}
for sphere, where $\Pi_\ud$ is
\begin{equation}\label{eq:djpd}
\Pi_{\ud}=\frac{1}{\epsilon^{2}}\frac{(n-1)(m-1)}{2b(n-m)}\theta_{\ue}^{2}[(\frac{b}{h})^{n}-(\frac{b}{h})^{m}]
\end{equation}
$\epsilon=H/R$ is an aspect ratio if the radius $R$ is used as the length scale, $b=h_{min}/H$ is dimensionless thickness of the stable film which is compatible with the precursor thickness.

The effect of disjoining pressure can be also introduced in the linear perturbation equation on cylindrical or spherical surface, if the expression of $\Pi_\ud$ is linearized
\begin{subequations}
\begin{equation}\label{eq:djpdod}
\Pi_{\ud}=\Pi_{\ud0}+\epsilon(\partial_{h_{0}}\Pi_{\ud0})G
\end{equation}
\begin{equation}\label{eq:djpd0}
\Pi_{\ud0}=\frac{1}{\epsilon^{2}}\frac{(n-1)(m-1)}{2b(n-m)}\theta_{\ue}^{2}[(\frac{b}{h_{0}})^{n}-(\frac{b}{h_{0}})^{m}]
\end{equation}
\begin{equation}\label{eq:djpdh}
\partial_{h_{0}}\Pi_{\ud0}=\frac{1}{\epsilon^{2}}\frac{(n-1)(m-1)}{2b(n-m)}\frac{\theta_{\ue}^{2}}{h_{0}}[m(\frac{b}{h_{0}})^{m}-n(\frac{b}{h_{0}})^{n}]
\end{equation}
\end{subequations}
Thus the linear perturbation equations become
\begin{subequations}\label{eq:cypertqdjp}
\begin{equation}
\frac{\partial G}{\partial t}+\frac{\partial}{\partial\theta}[-h_{0}^{3}\frac{\partial p_{\ug}}{\partial \theta}+3h_{0}^{2}(-\frac{\partial p_{s}}{\partial\theta}+\sin\theta)G]+h_{0}^{3}q^{2}p_{\ug}=0
\end{equation}
\begin{equation}\label{eq:cypsdjp}
p_{s}=-\frac{1}{Bo}(h_{0}+\frac{\partial^{2}h_{0}}{\partial\theta^{2}}+\Pi_{\ud0})
\end{equation}
\begin{equation}\label{eq:cypgqdjp}
p_{\ug}=-\frac{1}{Bo}(G+\frac{\partial^{2}G}{\partial\theta^{2}}-q^{2}G+\partial_{h_{0}}\Pi_{\ud0}G)
\end{equation}
\end{subequations}
for cylindrical problem, and
\begin{subequations}\label{eq:sppertqdjp}
\begin{equation}
\frac{\partial G}{\partial t}+\frac{1}{\sin\theta}\frac{\partial}{\partial \theta}{\sin\theta[-h_{0}^{3}\frac{\partial p_{\ug}}{\partial\theta}+3h_{0}^{2}(-\frac{\partial p_{s}}{\partial\theta}+\sin\theta)G]}+\frac{h_{0}^{3}q^{2}p_{\ug}}{\sin^{2}\theta}=0
\end{equation}
\begin{equation}\label{eq:sppsdjp}
p_{s}=-\frac{1}{Bo}(2h_{0}+\frac{1}{\tan\theta}\frac{\partial h_{0}}{\partial\theta}+\frac{\partial^{2}h_{0}}{\partial\theta^{2}}+\Pi_{\ud0})
\end{equation}
\begin{equation}\label{eq:sppgqdjp}
p_{\ug}=-\frac{1}{Bo}(2G+\frac{1}{\tan\theta}\frac{\partial G}{\partial\theta}+\frac{\partial^{2}G}{\partial \theta^{2}}-\frac{q^{2}G}{\sin^{2}\theta}+
\partial_{h_{0}}\Pi_{\ud0}G)
\end{equation}
\end{subequations}
for spherical problem.

For complete wetting cases, the equilibrium contact angles $\theta_{\ue}$ is zero and the disjoining pressure can be completely neglected. The forms of capillary wave and linear perturbation equations will not change.

\section{\label{sec:ast}Asymptotic theory}

For the liquid films which have the same physical property, the high $Bo$ represents thinner film flow on larger sized substrate. A method of matched asymptotic expansions had been used to analyze the two-dimensional and axisymmetric evolution equations (\ref{eq:tdeq}) and (\ref{eq:axeq}), if $Bo$ in these equations is sufficiently large \citep{Hou2017}. For high $Bo$ flow, the two-dimensional or axisymmetric capillary wave always consists of three regions - outer, inner and contact region, each with its own characteristic scaling. We have proved that there are inherent similarities in inner region for the cylindrical and spherical problems \citep{Hou2017}. According to the deduction of the asymptotic theory, the leading order inner equation on cylindrical surface is identical to that on spherical surface, and the effect of curved substrate can be considered as higher order corrections. This feature can be demonstrated using an asymptotic behavior in the inner region because the morphology of the capillary ridge are closer to the profile calculated using the asymptotic theory as $Bo$ increases.

The concepts in the asymptotic theory which is used to treat capillary waves can be extended to analyze the high $Bo$ limit of the linear perturbation equations. There is no limitation for the extent of perturbation function $G$, which implies that it can cover the entire substrate. However, if we only pay attention to the localized disturbances near the moving front, which has been proved to be the most dominant factor of the fingering instability,\citep{Davis2003} the solution of linear perturbation equations may be confined in the inner region. In the following subsections, we will transform the original linear perturbation equations into the inner system and derive leading order modal equation which can be considered as an eigenvalue problem.

\subsection{Cylindrical perturbation equation in inner region}

We define an inner coordinate system moving with the front of capillary wave
\begin{equation}\label{eq:innerco}
\xi=Bo^{\frac{1}{3}}[\theta-\theta_{\uF}(t)]
\end{equation}
where $\theta_\uF$ is the moving front location. By substituting (\ref{eq:innerco}) the linear perturbation equation (\ref{eq:cyperta}) can be transformed in the inner system
\begin{subequations}
\begin{multline}\label{eq:cypertqin}
Bo^{-1/3}\frac{\partial G}{\partial t}-\dot{\theta}_{\uF}\frac{\partial G}{\partial\xi}+\frac{\partial}{\partial\xi}\{-Bo^{1/3}h_{0}^{3}
\frac{\partial p_{\ug}}{\partial\xi}+3h_{0}^{2}[-Bo^{1/3}\frac{\partial p_{s}}{\partial\xi}+\\
\sin(\theta_{\uF}+Bo^{-1/3}\xi)]G\}-Bo^{-1/3}h_{0}^{3}q^{2}p_{\ug}=0
\end{multline}
\begin{equation}\label{eq:cypsin}
p_{s}=-Bo^{-1}h_{0}-Bo^{-1/3}\frac{\partial^{2}h_{0}}{\partial\xi^{2}}
\end{equation}
\begin{equation}\label{eq:cypgqin}
p_{\ug}=-Bo^{-1}G-Bo^{-1/3}\frac{\partial^{2}G}{\partial\xi^{2}}+Bo^{-1}q^{2}G
\end{equation}
\end{subequations}
in which the front speed $\dot{\theta}_\uF$ can be calculated using a Rankine-Hugoniot relation \citep{Hou2017}
\begin{equation}\label{eq:dthetaf}
\dot{\theta}_{\uF}=\sin\theta_{\uF}(h_{\uF}^{2}+h_{\uF}b+b^{2})
\end{equation}
where $h_\uF$ is the front thickness at the front location.

When $Bo^{1/3}\gg1$, by neglecting the $O(Bo^{-1/3})$ and $O(Bo^{-2/3})$ terms, the leading order of (\ref{eq:cypertqin}) is
\begin{multline}\label{eq:cypertqloin}
Bo^{-1/3}\frac{\partial G}{\partial t}=\dot{\theta}_{\uF}\frac{\partial G}{\partial\xi}-\frac{\partial}{\partial\xi}\{h_{0}^{3}
[\frac{\partial^{3}G}{\partial\xi^{3}}-(Bo^{-1/3}q)^{2}\frac{\partial G}{\partial\xi}]+
3h_{0}^{2}(\frac{\partial^{3}h_{0}}{\partial\xi^{3}}+\sin\theta_{\uF})G\}+\\
h_{0}^{3}[(Bo^{-1/3}q)^{2}\frac{\partial^{2}G}{\partial\xi^{2}}-(Bo^{-1/3}q)^{4}G]
\end{multline}
The reason of retaining the terms including coefficient $Bo^{-1/3}$ in (\ref{eq:cypertqloin}) will be explained below. We continue to use
\begin{subequations}
\begin{equation}\label{eq:innerh}
h=h_{\uF}h', G=h_{\uF}G'
\end{equation}
\begin{equation}\label{eq:innerxi}
\xi=\frac{h_{\uF}^{1/3}}{\sin^{1/3}\theta_{\uF}}\xi'
\end{equation}
\end{subequations}
to transform (\ref{eq:cypertqloin}) into the following
\begin{multline}\label{eq:cypertqinn}
\frac{1}{h_{\uF}^{8/3}\sin^{4/3}\theta_{\uF}Bo^{1/3}}\frac{\partial G'}{\partial t}=(1+\delta+\delta^{2})\frac{\partial G'}{\partial\xi'}-
\frac{\partial}{\partial\xi'}\{h_{0}'^{3}[\frac{\partial^{3}G'}{\partial\xi'^{3}}-
(\frac{qh_{\uF}^{1/3}}{\sin^{1/3}\theta_{\uF}Bo^{1/3}})^{2}\frac{\partial G'}{\partial\xi'}]+\\ 3h_{0}'^{2}(\frac{\partial^{3}h'_{0}}{\partial\xi'^{3}}+1)G'\}+
h_{0}'^{3}[(\frac{qh_{\uF}^{1/3}}{\sin^{1/3}\theta_{\uF}Bo^{1/3}})^{2}\frac{\partial^{2}G'}{\partial\xi'^{2}}-
(\frac{qh_{\uF}^{1/3}}{\sin^{1/3}\theta_{\uF}Bo^{1/3}})^{4}G']
\end{multline}
where $\delta(t)=b/h_\uF(t)$ can be considered as a relative precursor thickness. As the base state of linear perturbation, the leading order inner profile satisfies \citep{Hou2017}
\begin{subequations}
\begin{equation}\label{eq:innode}
\frac{\partial^{3}h_0'}{\partial\xi'^{3}}=\frac{1+\delta+\delta^{2}}{h_0'^{2}}-\frac{\delta+\delta^{2}}{h_0'^{3}}-1
\end{equation}
\begin{equation}\label{eq:innerbc1}
h_0'\rightarrow 1 \quad \xi'\rightarrow -\infty
\end{equation}
\begin{equation}\label{eq:innerbc2}
h_0'\rightarrow \delta \quad \xi'\rightarrow +\infty
\end{equation}
\end{subequations}
The quasi-steady capillary wave $h'_0[\xi',\delta(t)]$ is dependent on time, so that the linear operator of $G'$ on the right side of (\ref{eq:cypertqinn}) includes explicit time-varying parts. Supposing we can consider that the time scale of perturbation evolution is much smaller than the time scale of base state evolution, mathematically, an inner time $\tau$ for linear perturbation can be defined as
\begin{equation}\label{eq:innert}
\tau=h_{\uF}^{8/3}\sin^{4/3}\theta_{\uF}Bo^{1/3}(t-t_{0})
\end{equation}
$t_0$ is current time. For $\Delta t=t-t_0$, we have $\Delta t/\tau\ll 1$ in the high $Bo$ flow. The base state can be rewritten as $h'_0[\xi',\delta(t_{0}+\tau/h_{\uF}^{8/3}\sin^{4/3}\theta_{\uF}Bo^{1/3})]$, where the time-varying parts can be neglected for high $Bo$ case. Moreover, if we only consider the sufficiently large wave numbers of $O(Bo^{1/3})$ in (\ref{eq:cypertqin}) and neglect all smaller terms, namely, we can define an inner wave number
\begin{equation}\label{eq:cyinnerq}
q'=\frac{qh_{\uF}^{1/3}}{\sin^{1/3}\theta_{\uF}Bo^{1/3}}
\end{equation}
in the inner region (the original $q$ can be called outer wave number). By substituting (\ref{eq:innert}) and (\ref{eq:cyinnerq}) into (\ref{eq:cypertqinn}), it can be rewritten as
\begin{multline}\label{eq:pertq}
\frac{\partial G'}{\partial\tau}=(1+\delta+\delta^{2})\frac{\partial G'}{\partial\xi'}-
\frac{\partial}{\partial\xi'}[h_{0}'^{3}(\frac{\partial^{3}G'}{\partial\xi'^{3}}-q'^{2}\frac{\partial G'}{\partial\xi'})+
3h_{0}'^{2}(\frac{\partial^{3}h'_{0}}{\partial\xi'^{3}}+1)G']+\\
h_{0}'^{3}(q'^{2}\frac{\partial^{2}G'}{\partial\xi'^{2}}-q'^{4}G')
\end{multline}
which is equivalent to the linear perturbation equation on a slope \citep{Bertozzi1997}. Equations (\ref{eq:innert}) and (\ref{eq:cyinnerq}) are two additional asymptotic conditions to simplify the inner perturbation equation in high $Bo$ limit. The evolution of perturbation in the inner region degenerates into the planar problem for a high $Bo$ case. Since the base state $h'_0$ does not depend on time within the time scale of perturbation evolution, a normal mode $G'(\xi',\beta)e^{\beta\tau}$ can replace $G'(\xi',\tau)$
\begin{multline}\label{eq:eigen}
\beta G'=(1+\delta+\delta^{2})\frac{\partial G'}{\partial \xi'}-
\frac{\partial}{\partial \xi'}[h_{0}'^{3}(\frac{\partial^{3}G'}{\partial\xi'^{3}}-q'^{2}\frac{\partial G'}{\partial \xi'})+
3h_{0}'^{2}(\frac{\partial^{3}h'_{0}}{\partial\xi'^{3}}+1)G']+\\
h_{0}'^{3}(q'^{2}\frac{\partial^{2}G'}{\partial\xi'^{2}}-q'^{4}G')
\end{multline}
which becomes an eigenvalue problem with decaying boundary conditions
\begin{subequations}\label{eq:eigenbc}
\begin{equation}\label{eq:eigenbc1}
G',G'_{\xi'}\rightarrow 0 \quad \xi'\rightarrow -\infty
\end{equation}
\begin{equation}\label{eq:eigenbc2}
G',G'_{\xi'}\rightarrow 0 \quad \xi'\rightarrow +\infty
\end{equation}
\end{subequations}
Note that this eigenvalue equation is also identical to that derived in the canonical problem concerning coating flow on a slope \citep{Spaid1996}. A dispersion relationship $\beta(q')$ for the largest eigenvalue versus the wave number can be calculated, and the most unstable inner wave number $q'_\umax$ is selected from the discrete point which is the peak of the $\beta(q')$ curve. With this result, the most unstable outer wave number $q_\umax$ is computed as
\begin{equation}\label{eq:cyouterqmax}
q_{\umax}=\frac{q'_{\umax}\sin^{1/3}\theta_{\uF}Bo^{1/3}}{h_{\uF}^{1/3}}
\end{equation}

Equation (\ref{eq:cyouterqmax}) can be applied to estimating the most unstable wave number or even the fingering number which evolves on cylindrical surface using the given values of $Bo$, $\theta_\uF$ and $h_\uF$ at a certain moment. This estimation is accurate to the leading order for high $Bo$ flow. It implies that there is a power law for the fingering number which may be formed in a coating flow on upper cylinder. The most unstable wave number is proportional to $Bo^{1/3}$ and $\sin^{1/3}\theta_\uF$, and inversely proportional to $h_\uF^{1/3}$. The front thickness $h_\uF$ is not independent and is related to front location $\theta_\uF$ according to the outer solution of (\ref{eq:tdeq}). Because the front location and front thickness vary with time, the fingering number is certainly dependent on the onset of fingering instability. Note that the time scale for perturbation in the inner region is different from the outer region, if the growth rate $\beta$ for the most unstable inner wave number is obtained from the eigenvalue of (\ref{eq:eigen}), the growth rate $\gamma$ in outer region can be computed using (\ref{eq:innert})
\begin{equation}\label{eq:outerrate}
\gamma=h_{\uF}^{8/3}\sin^{4/3}\theta_{\uF}Bo^{1/3}\beta
\end{equation}
because the dimension of growth rate is the reciprocal of time scale. In Sec.~\ref{sec:resdis} we will show that the growth rate $\beta$ increases significantly as the parameter $\delta$ decreases. According to (\ref{eq:outerrate}), the growth rate $\gamma$ in outer region is influenced by front location $\theta_\uF$, Bond number $Bo$ and precursor thickness $b$.

\subsection{Spherical perturbation equation in inner region}

For the spherical problem, the same transformation (\ref{eq:innerco}) can be used to obtain the expressions of linear perturbation equation (\ref{eq:spperta}) in the inner system
\begin{subequations}
\begin{multline}\label{eq:sppertqin}
Bo^{-1/3}\frac{\partial G}{\partial t}-\dot{\theta}_{\uF}\frac{\partial G}{\partial\xi}+\frac{1}{\sin(\theta_{\uF}+Bo^{-1/3}\xi)}\frac{\partial}{\partial\xi}\{-\sin(\theta_{\uF}+Bo^{-1/3}\xi)
Bo^{1/3}h_{0}^{3}\frac{\partial p_{\ug}}{\partial\xi}+\\
3\sin(\theta_{\uF}+Bo^{-1/3}\xi)h_{0}^{2}[-Bo^{1/3}\frac{\partial p_{s}}{\partial\xi}+ \sin(\theta_{\uF}+Bo^{-1/3}\xi)]G\}+\\
\frac{Bo^{-1/3}h_{0}^{3}q^{2}p_{\ug}}{\sin^{2}(\theta_{\uF}+Bo^{-1/3}\xi)}=0
\end{multline}
\begin{equation}\label{eq:sppsin}
p_{s}=-2Bo^{-1}h_{0}-\frac{Bo^{-2/3}}{\tan(\theta_{\uF}+Bo^{-1/3}\xi)}\frac{\partial h_{0}}{\partial\xi}-Bo^{-1/3}\frac{\partial^{2}h_{0}}{\partial\xi^{2}}
\end{equation}
\begin{equation}\label{eq:sppgqin}
p_{\ug}=-2Bo^{-1}G-\frac{Bo^{-2/3}}{\tan(\theta_{\uF}+Bo^{-1/3}\xi)}\frac{\partial G}{\partial\xi}-Bo^{-1/3}\frac{\partial^{2}G}{\partial\xi^{2}}+
\frac{Bo^{-1}q^{2}}{\sin^{2}(\theta_{\uF}+Bo^{-1/3}\xi)}G
\end{equation}
\end{subequations}
The leading order of (\ref{eq:sppertqin}) is
\begin{multline}\label{eq:sppertqloin}
\frac{1}{Bo^{1/3}}\frac{\partial G}{\partial t}=\dot{\theta}_{\uF}\frac{\partial G}{\partial\xi}-\frac{\partial}{\partial\xi}\{h_{0}^{3}
[\frac{\partial^{3}G}{\partial\xi^{3}}-(\frac{q}{Bo^{1/3}})^{2}\frac{1}{\sin^{2}\theta_{\uF}}\frac{\partial G}{\partial\xi}]+
3h_{0}^{2}(\frac{\partial^{3}h_{0}}{\partial\xi^{3}}+\sin\theta_{\uF})G\}+\\
\frac{h_{0}^{3}}{\sin^{2}\theta_{\uF}}[(\frac{q}{Bo^{1/3}})^{2}\frac{\partial^{2}G}{\partial\xi^{2}}-(\frac{q}{Bo^{1/3}})^{4}\frac{G}{\sin^{2}\theta_{\uF}}]
\end{multline}
This equation can be transformed further using (\ref{eq:innerh}) and (\ref{eq:innerxi}) (the front speed $\dot{\theta}_\uF$ is identical to (\ref{eq:dthetaf})) into the following
\begin{multline}\label{eq:sppertqinn}
\frac{1}{h_{\uF}^{8/3}\sin^{4/3}\theta_{\uF}Bo^{1/3}}\frac{\partial G'}{\partial t}=(1+\delta+\delta^{2})\frac{\partial G'}{\partial\xi'}-
\frac{\partial}{\partial\xi'}\{h_{0}'^{3}[\frac{\partial^{3}G'}{\partial\xi'^{3}}-
(\frac{qh_{\uF}^{1/3}}{\sin^{4/3}\theta_{\uF}Bo^{1/3}})^{2}\frac{\partial G'}{\partial\xi'}]+\\ 3h_{0}'^{2}(\frac{\partial^{3}h'_{0}}{\partial\xi'^{3}}+1)G'\}+
h_{0}'^{3}[(\frac{qh_{\uF}^{1/3}}{\sin^{4/3}\theta_{\uF}Bo^{1/3}})^{2}
\frac{\partial^{2}G'}{\partial\xi'^{2}}-(\frac{qh_{\uF}^{1/3}}{\sin^{4/3}\theta_{\uF}Bo^{1/3}})^{4}G']
\end{multline}
If the time scale of linear perturbation evolution is assumed to be identical to the cylindrical problem (\ref{eq:innert}) and an inner wave number on spherical surface is defined as
\begin{equation}\label{eq:spinnerq}
q'=\frac{qh_{\uF}^{1/3}}{\sin^{4/3}\theta_{\uF}Bo^{1/3}}
\end{equation}
(\ref{eq:sppertqinn}) can be rewritten as
\begin{multline*}
\frac{\partial G'}{\partial\tau}=(1+\delta+\delta^{2})\frac{\partial G'}{\partial\xi'}-
\frac{\partial}{\partial\xi'}[h_{0}'^{3}(\frac{\partial^{3}G'}{\partial\xi'^{3}}-q'^{2}\frac{\partial G'}{\partial\xi'})+
3h_{0}'^{2}(\frac{\partial^{3}h'_{0}}{\partial\xi'^{3}}+1)G']+\\
h_{0}'^{3}(q'^{2}\frac{\partial^{2}G'}{\partial\xi'^{2}}-q'^{4}G')
\end{multline*}
which also degenerates into the eigenvalue problem (\ref{eq:eigen}) with boundary conditions (\ref{eq:eigenbc}). Similarly, the derived formula concerning the most unstable outer wave number $q_\umax$ is
\begin{equation}\label{eq:spouterqmax}
q_{\umax}=\frac{q'_{\umax}\sin^{4/3}\theta_{\uF}Bo^{1/3}}{h_{\uF}^{1/3}}
\end{equation}
Because the difference in definitions of the inner wave number between cylindrical problem (\ref{eq:cyinnerq}) and spherical problem (\ref{eq:spinnerq}), compared to formula (\ref{eq:cyouterqmax}) on cylindrical surface, there is an additional factor $\sin\theta_\uF$ in formula (\ref{eq:spouterqmax}) on spherical surface. The most unstable wave number is proportional to 4/3-th power of $\sin\theta_\uF$, which is the only distinction compared to 1/3-th power of $\sin\theta_\uF$ in cylindrical problem. Note that although the wave number of Fourier mode on spherical surface should be integer according to periodic condition (\ref{eq:pedbc}), (\ref{eq:spouterqmax}) can be still applied to estimating an approximate fingering number using a rounding-off method. The outer growth rate $\gamma$ calculated from the eigenvalue is identical to the cylindrical problem (\ref{eq:outerrate}), due to the same definition of the time scale.

\subsection{Disjoining pressure in inner region}

If the disjoining pressure is added into the base state and linear perturbation equations, the asymptotic theory under high $Bo$ condition can be extended to include the effect of partial wetting. It is known that the disjoining pressure can be neglected in the outer region where the film thickness $h\gg b$ and (\ref{eq:djpd}) becomes vanishingly small. The effect of disjoining pressure is only operative in the immediate vicinity of apparent contact line. Thus the form of linear perturbation equation in the inner region is more appropriate to introduce disjoining pressure terms. For cylindrical problem, (\ref{eq:cypertqdjp}) can be transformed in the inner system using (\ref{eq:innerco})
\begin{subequations}\label{eq:cypertqdjpin}
\begin{multline}
Bo^{-1/3}\frac{\partial G}{\partial t}-\dot{\theta}_{\uF}\frac{\partial G}{\partial\xi}+\frac{\partial}{\partial\xi}\{-Bo^{1/3}h_{0}^{3}
\frac{\partial p_{\ug}}{\partial\xi}+3h_{0}^{2}[-Bo^{1/3}\frac{\partial p_{s}}{\partial\xi}+\sin(\theta_{\uF}+Bo^{-1/3}\xi)]G\}-\\
Bo^{-1/3}h_{0}^{3}q^{2}p_{\ug}=0
\end{multline}
\begin{equation}\label{eq:cypsdjpin}
p_{s}=-Bo^{-1}h_{0}-Bo^{-1/3}\frac{\partial^{2}h_{0}}{\partial\xi^{2}}-Bo^{-1}\Pi_{\ud0}
\end{equation}
\begin{equation}\label{eq:cypgqdjpin}
p_{\ug}=-Bo^{-1}G-Bo^{-1/3}\frac{\partial^{2}G}{\partial\xi^{2}}+Bo^{-1}q^{2}G-Bo^{-1}\partial_{h_{0}}\Pi_{\ud0}G
\end{equation}
\end{subequations}
The leading order of (\ref{eq:cypertqdjpin}) is
\begin{multline}\label{eq:cypertqdjploin}
Bo^{-1/3}\frac{\partial G}{\partial t}=\dot{\theta}_{\uF}\frac{\partial G}{\partial\xi}-\frac{\partial}{\partial\xi}\{h_{0}^{3}
[\frac{\partial^{3}G}{\partial\xi^{3}}-(Bo^{-1/3}q)^{2}\frac{\partial G}{\partial\xi}+
Bo^{-2/3}\frac{\partial(\partial_{h_{0}}\Pi_{\ud0}G)}{\partial\xi}]+\\
3h_{0}^{2}(\frac{\partial^{3}h_{0}}{\partial\xi^{3}}+Bo^{-2/3}\frac{\partial\Pi_{\ud0}}{\partial\xi}+\sin\theta_{\uF})G\}+
h_{0}^{3}[(Bo^{-1/3}q)^{2}\frac{\partial^{2}G}{\partial\xi^{2}}-\\
(Bo^{-1/3}q)^{4}G+(Bo^{-1/3}q)^{2}Bo^{-2/3}\partial_{h_{0}}\Pi_{\ud0}G]
\end{multline}
Note that there is a coefficient $\epsilon^{-2}$ in the expressions of $\Pi_{\ud0}$ and $\partial_{h_0}\Pi_{\ud0}$ according to (\ref{eq:djpd0}) and (\ref{eq:djpdh}), if we assume that $Bo^{1/3}\epsilon\sim 1$, the terms including $(Bo^{1/3}\epsilon)^{-2}$ can be retained in the leading order linear perturbation equation in the inner region. Finally the quasi-steady eigenvalues equation become
\begin{multline}\label{eq:eigendjp}
\beta G'=(1+\delta+\delta^{2})\frac{\partial G'}{\partial \xi'}-
\frac{\partial}{\partial \xi'}\{h_{0}'^{3}[\frac{\partial^{3}G'}{\partial\xi'^{3}}-q'^{2}\frac{\partial G'}{\partial \xi'}+K\frac{\partial(\partial_{h_{0}}\Pi_{\ud0}'G')}{\partial\xi'}]+
3h_{0}'^{2}(\frac{\partial^{3}h'_{0}}{\partial\xi'^{3}}+1+\\
K\frac{\partial\Pi_{\ud0}'}{\partial\xi'})G'\}+h_{0}'^{3}(q'^{2}\frac{\partial^{2}G'}{\partial\xi'^{2}}-q'^{4}G'+q'^{2}K\partial_{h_{0}}\Pi_{\ud0}'G')
\end{multline}
where
\begin{subequations}
\begin{equation}\label{eq:djpcoin}
K=\frac{1}{(Bo^{1/3}\epsilon)^{2}h_{\uF}^{4/3}\sin^{2/3}\theta_{\uF}}\frac{(n-1)(m-1)\theta_{\ue}^{2}}{2\delta(n-m)}=(3Ca)^{-2/3}\frac{(n-1)(m-1)\theta_{\ue}^{2}}{2\delta(n-m)}
\end{equation}
\begin{equation}\label{eq:djpd0in}
\Pi_{\ud0}'=(\frac{\delta}{h_{0}'})^{n}-(\frac{\delta}{h_{0}'})^{m}
\end{equation}
\begin{equation}\label{eq:djpdhin}
\partial_{h_{0}}\Pi_{\ud0}'=\frac{1}{h_{0}'}[m(\frac{\delta}{h_{0}'})^{m}-n(\frac{\delta}{h_{0}'})^{n}]
\end{equation}
\end{subequations}
$K$ is called the dimensionless contact angle parameter, and $Ca=\epsilon^{3}Bo\sin\theta_{\uF}h_\uF^2/3$ is a capillary number \citep{Eres2000}. The base state $h_{0}'$ satisfies a equation which is identical to (\ref{eq:innode}) with the addition of disjoining pressure terms \citep{Hou2017}
\begin{equation}\label{eq:inndjpode}
\frac{\partial^{3}h_0'}{\partial\xi'^{3}}=\frac{1+\delta+\delta^{2}}{h_0'^{2}}-\frac{\delta+\delta^{2}}{h_0'^{3}}-1-
\frac{K}{h_0'}[m(\frac{\delta}{h_0'})^{m}-n(\frac{\delta}{h_0'})^{n}]\frac{\partial h_0'}{\partial\xi'}
\end{equation}
and the same boundary condition (\ref{eq:innerbc1}) and (\ref{eq:innerbc2}). Equation (\ref{eq:eigendjp}) can be considered as the combination of the original eigenvalues problem proposed by \citet{Spaid1996} with the effect of disjoining pressure. The identical leading order linear perturbation equation for spherical problem can be obtained from the transformation of (\ref{eq:sppertqdjp})in the inner region. Even though the disjoining pressure is considered, the high $Bo$ limits of the linear perturbation equations on both cylindrical and spherical surfaces degenerate into a common form. Note that the formulae (\ref{eq:cyouterqmax}) and (\ref{eq:spouterqmax}) will not change when (\ref{eq:eigendjp}) is used as the eigenvalues equation, because the transformation between the outer and inner wave number is not affected by the addition of disjoining pressure.

\section{\label{sec:num}Numerical approach}

It is important to construct a dispersion relationship $\beta(q')$ for extracting the most unstable inner wave number $q_\umax'$. The eigenvalue equation (\ref{eq:eigen}) should be solved to calculate the largest eigenvalue and corresponding eigenfunction for a given $q'$. An Evans function approach is widely used for eigenvalue problems in solitary waves and coating flow problems.\citep{Chang1998} This approach is essentially a numerical shooting method in which the profile of eigenfunction can be numerically integrated from a sufficiently far upstream point where the base state is uniform and the perturbation is very small. The numerical integration can be performed using a high order adaptive Runge-Kutta solver and the eigenvalue is naturally determined in the shooting process using the zeros of an Evans function, because the decaying boundary condition downstream should be satisfied. For eigenvalue problem (\ref{eq:eigendjp}) which includes the disjoining pressure effect, only three lower-order terms are added compared to (\ref{eq:eigen}) so that the Evans function method can be extended simply. The details concerning the Evans function shooting method for solving eigenvalue problem (\ref{eq:eigen}) and (\ref{eq:eigendjp}) can be found in the Appendix.

We have known that the modal analysis is only valid in an asymptotic theory if $Bo$ is sufficiently high. It is necessary to develop a numerical approach to solve the complete linear perturbation equations. The profile of linear perturbation at arbitrary time can be numerically solved from (\ref{eq:cyperta}) or (\ref{eq:spperta}) (in addition, (\ref{eq:cypertqdjp}) or (\ref{eq:sppertqdjp}) in partial wetting cases). The numerical method used here belongs to a time-marching finite difference method. A Crank-Nicolson scheme is used as the temporal discretization and the standard central difference scheme is applied to discretizing the linear operator, so that the second order accuracy for both time and space can be achieved \citep{Thomas1995}. Note that the linear perturbation equation should be solved simultaneously with the two-dimensional or axisymmetric capillary wave equation ((\ref{eq:tdeq}) or (\ref{eq:axeq})) because the variable coefficients concerning the base state in the linear perturbation equation should be updated as time increases. For each $q$, a typical initial condition is started, and the linear growth of the perturbation in an appropriate duration is examined. The integration of the perturbation profile can be used to extract the exponential growth rate at a certain time, which is important in transient growth analysis. The time marching continues until some terminal conditions are satisfied. The only complication for the numerical schemes is that it is necessary to use an adaptive mesh with refinement at the apparent contact line in order to completely resolve the profile \citep{Hou2017}. The detailed description for this numerical method is given in the Appendix.

\section{\label{sec:resdis}Results and discussion}

\subsection{Modal analysis}

\begin{figure}
  \includegraphics[width=0.49\textwidth]{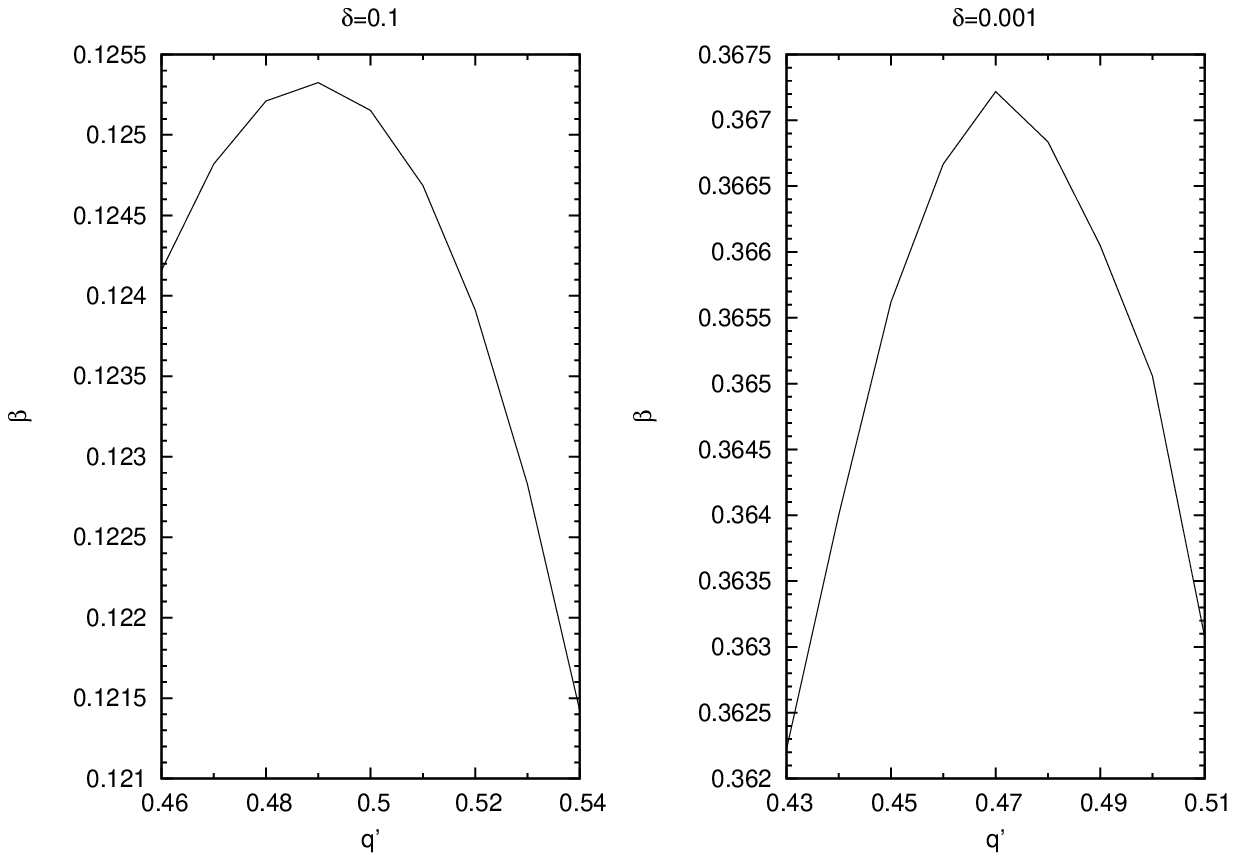}
  \includegraphics[width=0.49\textwidth]{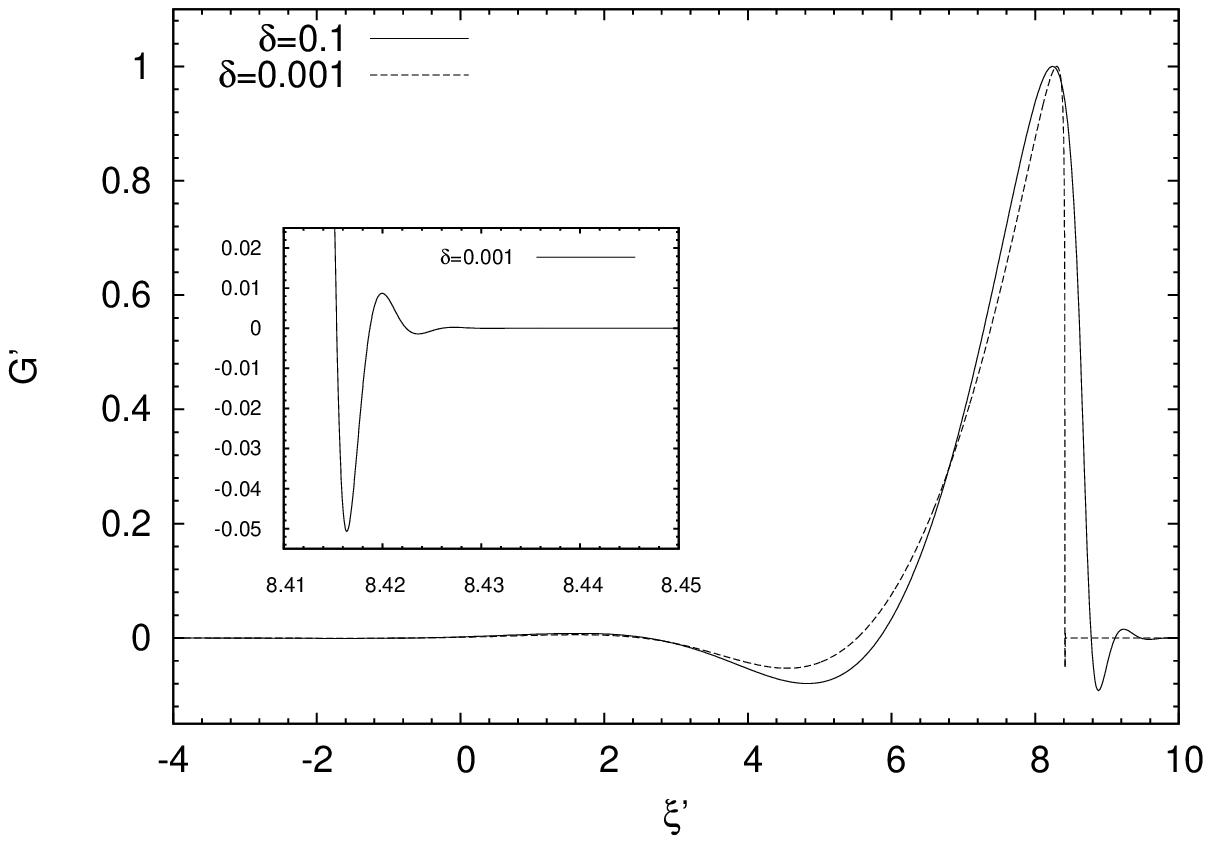}
  \centerline{(a)\hspace{0.45\textwidth}(b)}
  \caption{\label{fig:modal} (a) Dispersion curve of the largest eigenvalue versus the inner wave number near the most unstable region, for two precursor parameters $\delta=0.1$ and $\delta=0.001$. (b) Comparison of the eigenfunction profiles corresponding to the largest eigenvalue and most unstable inner wave number between the $\delta=0.1$ case and the $\delta=0.001$ case, the inset is the refined structure near the apparent contact line for $\delta=0.001$ case.}
\end{figure}

Figure~\ref{fig:modal}(a) shows the curves of dispersion relationship $\beta(q')$ near the most unstable wave number for two precursor layer conditions $\delta=0.1$ and $\delta=0.001$. For a given $q'$, the largest eigenvalue $\beta$ is calculated using the Evans function method, and the step size of $q'$ is 0.01. The most unstable inner wave number appears at $q'=0.49$ for $\delta=0.1$ case and $q'=0.47$ for $\delta=0.001$ case, red-shifts only 0.02 compared to $\delta=0.1$ case. The $\delta=0.001$ case can simulate a typical complete wetting case in which a relatively thinner precursor layer is used to model microscopic scale near the contact line. Thus keeping the accuracy to the first decimal place for a large range of $\delta$, the most unstable inner wave number occurs at approximately $q'=0.5$, which accords with the results given by \citet{Spaid1996}. The value of $q'_\umax$ is of $O(1)$ and therefore verifies the reasonability of scale defined in (\ref{eq:cyinnerq}). With the results of most unstable inner wave number, the most unstable outer (original) wave number is
\begin{equation}\label{eq:cymodalqmax}
q_{\umax}=\frac{0.5\sin^{1/3}\theta_{\uF}Bo^{1/3}}{h_{\uF}^{1/3}}
\end{equation}
for cylindrical problem and
\begin{equation}\label{eq:spmodalqmax}
q_{\umax}=\frac{0.5\sin^{4/3}\theta_{\uF}Bo^{1/3}}{h_{\uF}^{1/3}}
\end{equation}
for spherical problem. These two formulae can be used for practice if $Bo$ and the onset of fingering instability (to determine $\theta_\uF$ and $h_\uF$) are given. Note that for a practical cylinder of length $l$, the formed fingering number can be estimated using (\ref{eq:cymodalqmax})
\begin{equation*}\label{eq:cyfinum}
N_{\mathrm{f}}=\mathrm{nint}(\frac{q_{\umax}l}{2\pi R})
\end{equation*}
and for spherical problem, the formed fingering number is identical to the nearest integer of (\ref{eq:spmodalqmax}) directly
\begin{equation*}\label{eq:spfinum}
N_{\mathrm{f}}=\mathrm{nint}(q_{\umax})
\end{equation*}
where $N_{\mathrm{f}}$ is the fingering number and $\mathrm{nint}(x)$ is the nearest integer function.

Because of the explicitly larger $\beta$ for $\delta=0.001$ case (about triple compared to $\delta=0.1$ case), according to (\ref{eq:outerrate}), it is believed that the linear growth rate in a wetting case is much greater than the prewetting case in which a macroscopic precursor layer exists ahead of the front. It implies that the front with a thinner precursor layer is more unstable. The profiles of eigenfunction $G'$ corresponding to the most amplified eigenvalue and most unstable wave number $q'_\umax$ are shown in Fig.~\ref{fig:modal}(b), for $\delta=0.1$ and $\delta=0.001$. The region of rapid variation in the eigenfunction corresponds to the region of rapid variation in the base state near the precursor film. As $\xi'$ increases or decreases, the eigenfunction is periodically decaying, which implies the localization of the eigenfunction in the inner region. Note that compared to the eigenfunction for $\delta=0.1$ case, there is a refined structure near the apparent contact line for $\delta=0.001$ case, which indicates a much smaller contact region in a wetting case, as shown in the inset of Fig.~\ref{fig:modal}(b).

\subsection{Transient growth analysis}

The moving front of the base state may be disturbed at arbitrary time, and the initial disturbance may be small perturbation or finite amplitude. For small perturbation, there are two stages in the evolution process: linear stage and nonlinear stage. For finite amplitude disturbances, the evolution enters into nonlinear stage directly \citep{Ye1999}. The small perturbation is a basic assumption for linearizing the complete governing equations (\ref{eq:cyndge}) and (\ref{eq:spndge}). The derived linear perturbation equation can be solved directly to study the behavior of transient growth. There is no high $Bo$ assumption in the original expressions of linear perturbation equations (\ref{eq:cypertq}) and (\ref{eq:sppertq}), so that the solutions can be used as a datum to verify the accuracy of the asymptotic theory and modal analysis. We define a logarithmic Bond number $logBo=\lg(Bo)$ for convenience of expression.

Because the transient growth is related to initial conditions of disturbance, a Gaussian profile at the capillary ridge is chosen as the initial perturbation for both (\ref{eq:cypertq}) and (\ref{eq:sppertq})
\begin{equation}\label{eq:initpert}
G_{\ui}(\theta)=Ae^{-(\frac{\theta-\theta_{\umax}}{\sigma})^{2}}
\end{equation}
where $A$ is the amplitude of initial perturbation, $\theta_\umax$ is peak location of the capillary ridge and $\sigma$ is the root mean square (RMS) width of perturbation function. This type of initial perturbation may be representative because the disturbances are possibly random along the $\theta$ direction in practice. The RMS width of initial perturbation is set to the width of the capillary ridge (defined using the distance between primary minimum and secondary minimum in the capillary wave profile). A sufficiently small value is chosen as the amplitude of initial perturbation to model an infinitesimal disturbance. The time when the initial perturbation is superimposing is the onset of fingering instability. For simplicity, before the onset, the capillary wave (base state) is assumed not to be disturbed by any disturbance, and after the onset, there is no new disturbance imposing. The transient growth rate can be defined after the profile of perturbation at a given time is constructed
\begin{equation}\label{eq:growthrate}
\gamma=\frac{\langle G,\frac{\partial G}{\partial t}\rangle}{\langle G,G\rangle}=\frac{\frac{1}{2}\frac{\partial}{\partial t}\langle G,G\rangle}{\langle G,G\rangle}=\frac{\langle G,L_{h_{0}}G\rangle}{\langle G,G\rangle}
\end{equation}
where $L_{h_0}$ is the complete linear operator in (\ref{eq:cypertq}) or (\ref{eq:sppertq}). The square 2-norm of perturbation profile is
\begin{equation*}
\langle G,G\rangle=\int G^{2}\ud\theta
\end{equation*}
for cylindrical problem, or
\begin{equation*}
\langle G,G\rangle=\int G^{2}\sin\theta\ud\theta
\end{equation*}
for spherical problem. Thus the transient growth rate can be considered as half the time derivatives of the square 2-norm. We can also define an integral growth rate to evaluate the average growth rate of the linear perturbation for a given wave number
\begin{equation}\label{eq:avrate}
\bar{\gamma}(q)=\frac{\int_{0}^{T}\gamma(q,t)\ud t}{T}
\end{equation}
The period $T$ is the duration from the onset of instability to the time when the linear evolution ends. The average growth rate $\bar{\gamma}(q)$ can be used to obtain a dispersion relationship which is similar to the dispersion curve $\beta(q')$ in the eigenvalue problem. The most unstable wave number is selected using the value that has the largest average growth rate and can be used as a contrast to the formulae (\ref{eq:cymodalqmax}) and (\ref{eq:spmodalqmax}) which is derived by modal analysis.

\begin{figure}
  \includegraphics[width=0.49\textwidth]{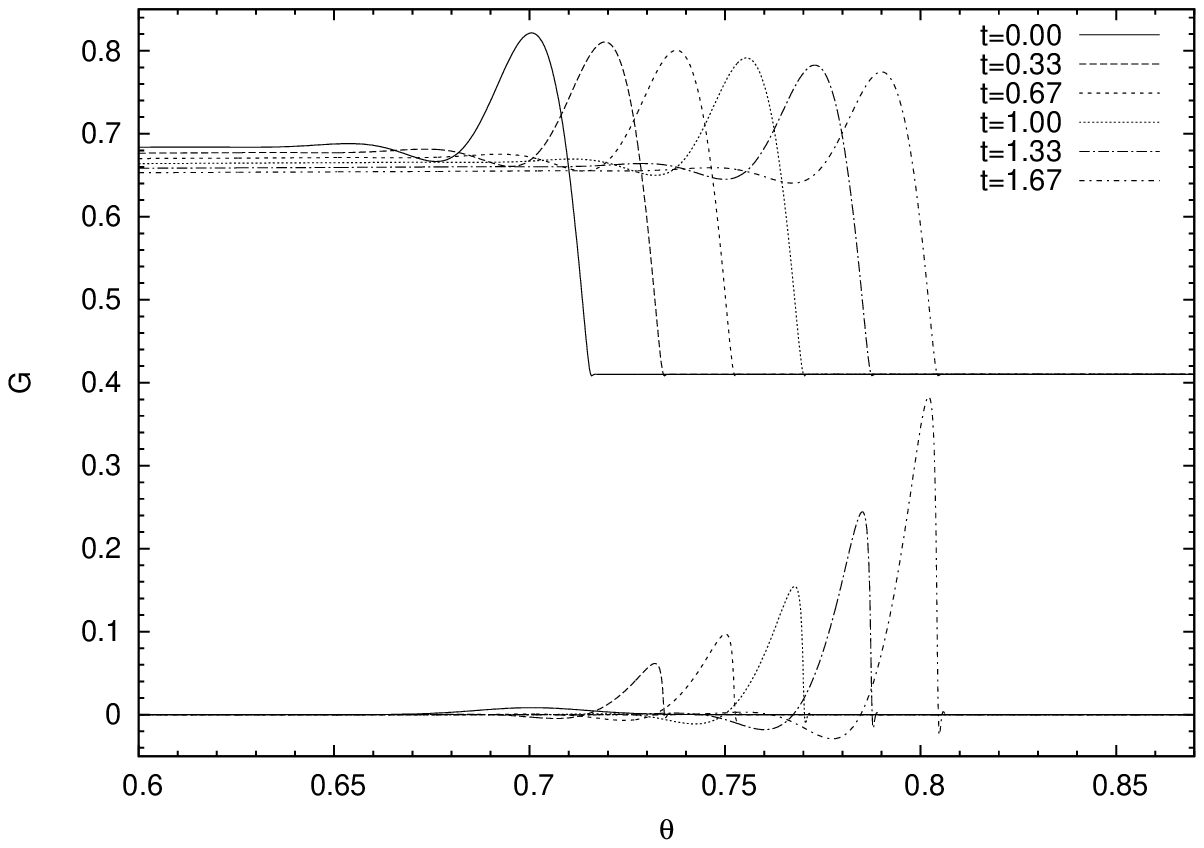}
  \includegraphics[width=0.49\textwidth]{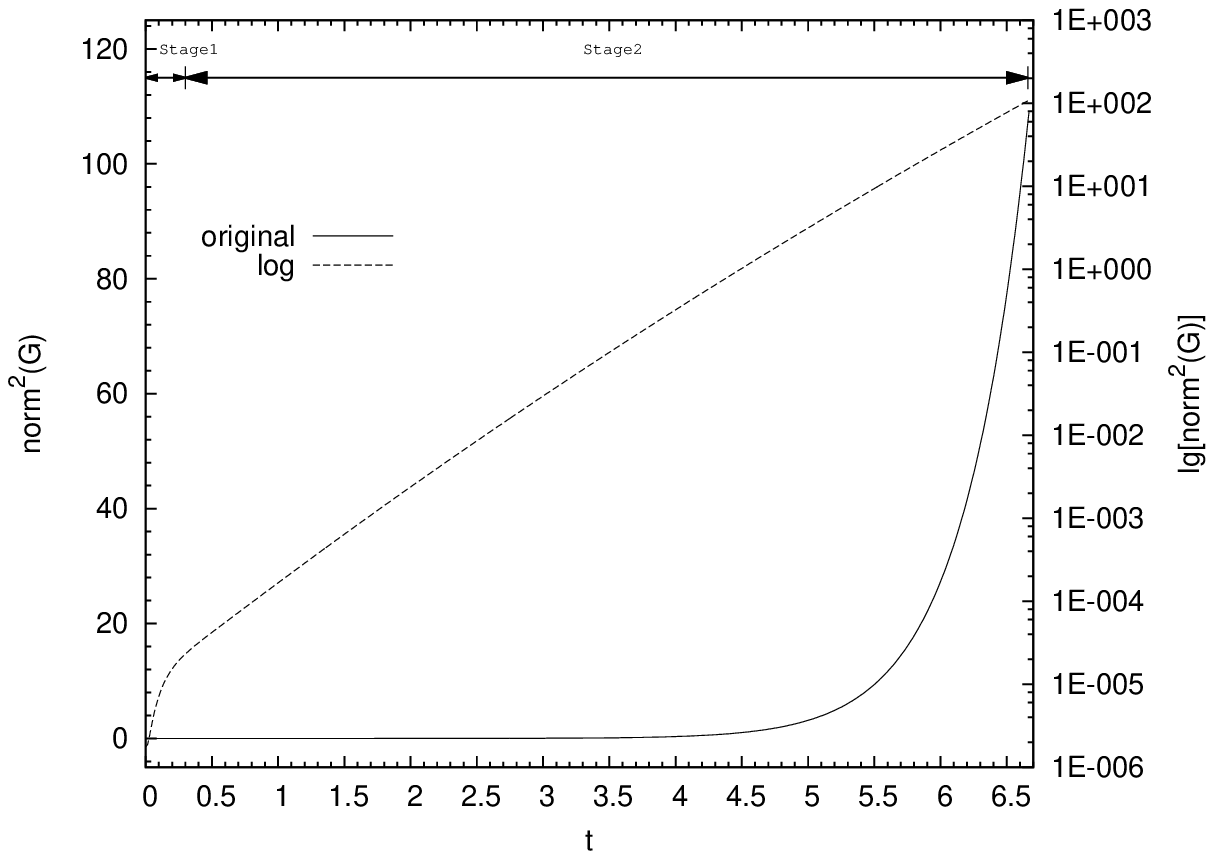}
  \centerline{(a)\hspace{0.45\textwidth}(b)}
  \caption{\label{fig:cyprofnormlb} (a) Transient profiles of linear perturbation and corresponding two-dimensional capillary wave on cylindrical surface for precursor thickness $b=0.01$. The parameters are $logBo=6.0$ and $q=68$. The initial perturbation is superimposed at $t=6$, timing from initial releasing moment. (b) Integral intensity of the linear perturbation as a function of time. The original value versus time is plotted as a solid line and the logarithmic value versus time is plotted as a dashed line.}
\end{figure}

For cylindrical problem, Fig.~\ref{fig:cyprofnormlb}(a) shows the transient profiles of linear perturbation and corresponding two-dimensional capillary wave at time ranging from $t=0$ to $t=5/3$. The Bond number is $logBo=6.0$ and precursor thickness is $b=0.01$. The onset of instability $t=0$ is corresponding to $t=6$, timing from initial moment when the film was releasing, and the wave number is $q=68$. For comparison purpose, the profiles of base-state capillary wave are translated to the upper part of the plot and zoomed out. The front of capillary wave is traveling at $\theta_\uF\approx$0.7-0.8 in the range of disturbance time. It is shown that the perturbation profile travels with the base state at an identical speed. The initial Gaussian perturbation is quickly changed into a different type profile at early time. This profile is similar to the eigenmode $G'$ solved from the eigenvalue problem (\ref{eq:eigen}) as shown in Fig.~\ref{fig:modal}(b). Then this quasi-eigenmode is amplified rapidly and the intensity of linear perturbation may exceed the thickness of capillary wave somewhere, the perturbation evolution should enter into the nonlinear stage. Figure \ref{fig:cyprofnormlb}(b) shows integral intensity of the perturbation profiles as a function of time. Both the original value versus time plot (solid line) and the logarithmic value versus time plot (dashed line) are shown. The integral intensity (square 2-norm) varies with time quickly, and the tendency of exponential growth is achieved finally. From the logarithmic plot, we can see that there are two main stages for linear perturbation evolution. In Stage 1 at early time, the slope (growth rate) is greater, which implies a faster growth. This early process can be considered as a transition from the Gaussian profile to the quasi-eigenmode, which is relative to the initial perturbation due to the nonnormal characteristic of the linear operator $L_{h_0}$. In Stage 2, the curve approximates a straight line. It indicates the eigenmode that has the largest eigenvalue dominates the linear perturbation evolution. But strictly speaking, the slope in Stage 2 is not a constant and in fact decreases slowly as time increases. The growth rate may vary with time, which is different from the modal analysis. The reason is that the time evolution of the base state (capillary wave) profile cannot be neglected if the linear perturbation evolution sustains for a long time. The front location and thickness have changed explicitly as shown in the upper part of Fig.~\ref{fig:cyprofnormlb}(a), and this change may introduce perturbation terms in linear operator $L_{h_0}$ and produce quasi-eigenmode of varying growth rate. From the asymptotic aspect according to (\ref{eq:outerrate}), the growth rate $\gamma$ is not large due to the relatively smaller $Bo$ ($logBo=6.0$) and thicker precursor thickness ($b=0.01$).

\begin{figure}
  \includegraphics[width=0.49\textwidth]{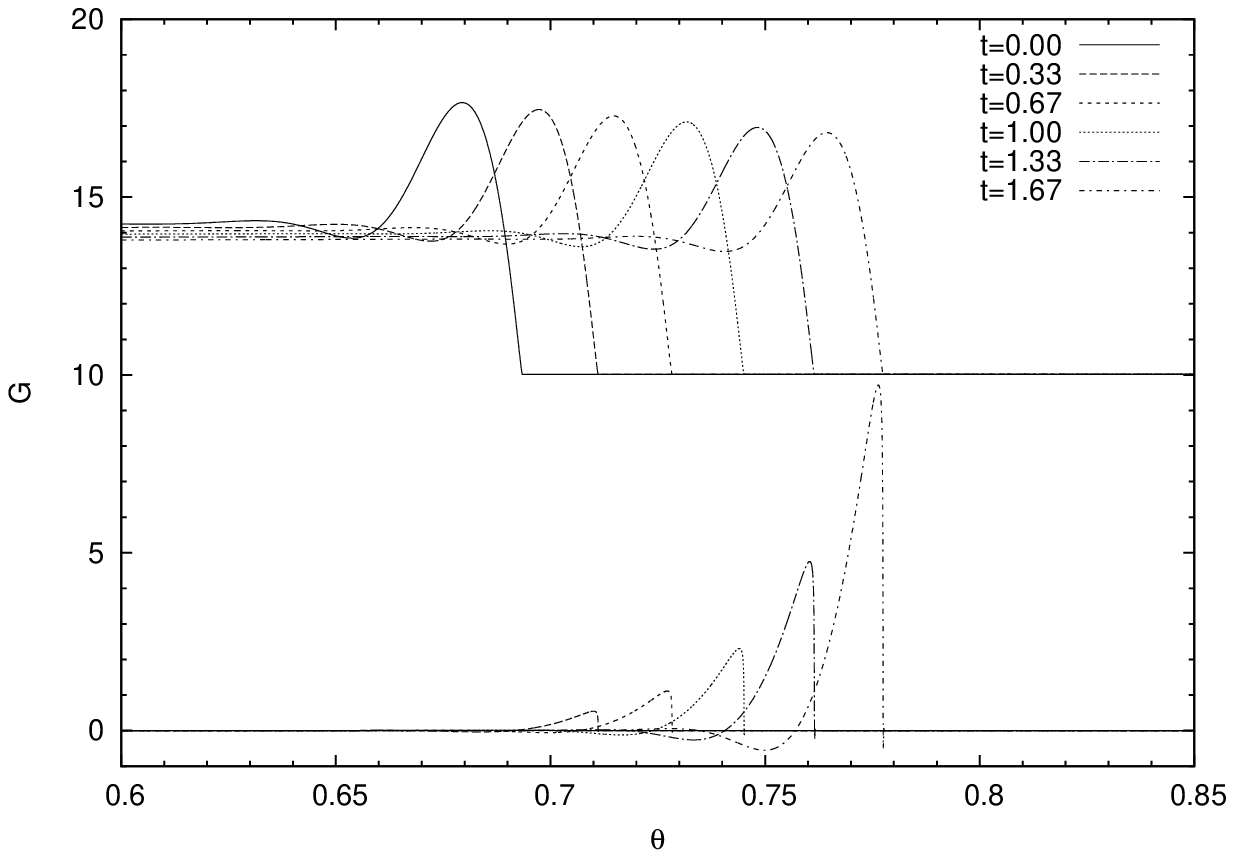}
  \includegraphics[width=0.49\textwidth]{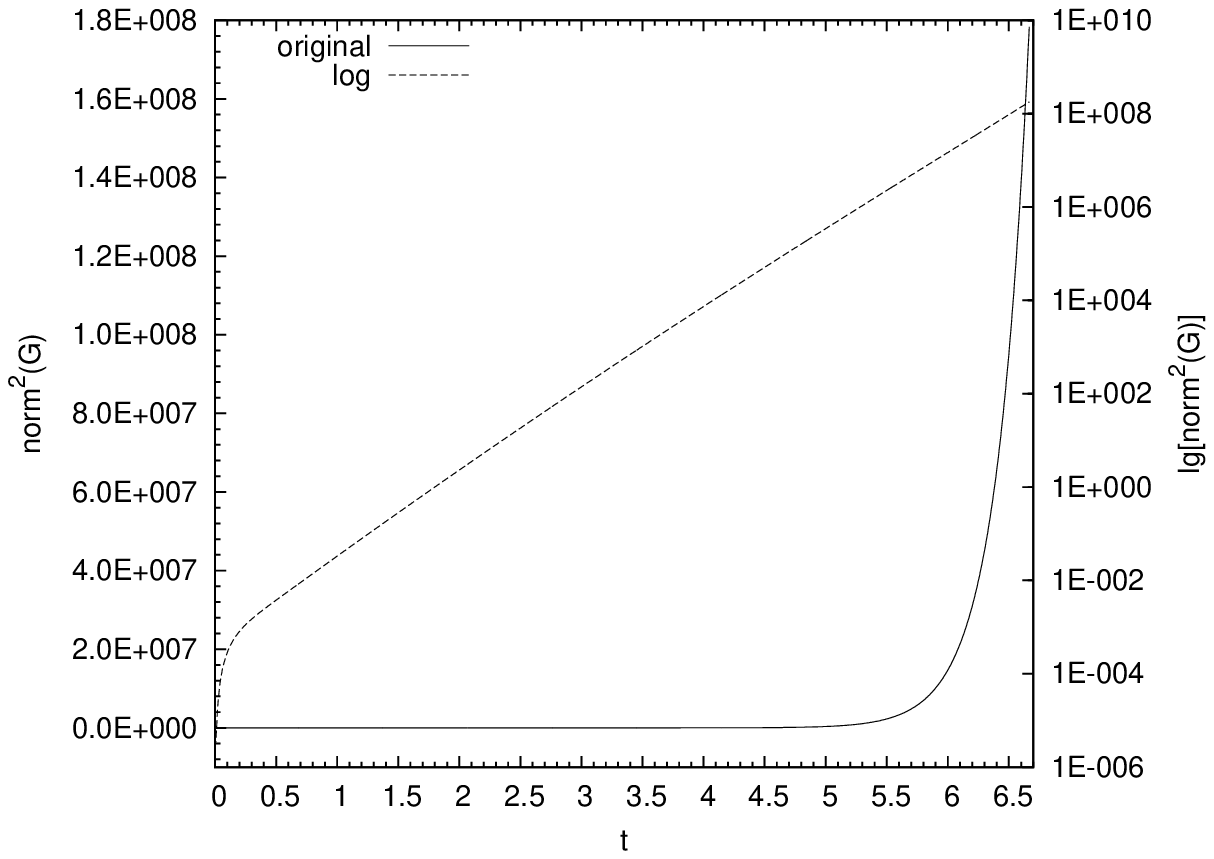}
  \centerline{(a)\hspace{0.45\textwidth}(b)}
  \caption{\label{fig:cyprofnormsb} (a) Transient profiles of linear perturbation and corresponding two-dimensional capillary wave on cylindrical surface for precursor thickness $b=0.001$. The parameters are identical to that in Fig.~\ref{fig:cyprofnormlb}(a). (b) The original and logarithmic integral intensity as a function of time.}
\end{figure}

To present the influence of a relatively thinner precursor layer on the linear growth, Fig.~\ref{fig:cyprofnormsb}(a) shows the profiles of linear perturbation and corresponding capillary wave for precursor thickness $b=0.001$. The range of time and the value of $Bo$ are identical to the $b=0.01$ case. The onset of instability corresponds to the two-dimensional capillary wave with a $b=0.001$ precursor layer at $t=6$. Compared to Fig.~\ref{fig:cyprofnormlb}(a), it is shown that the change of base state profile is slight, and the intensity of perturbation profile grows more quickly than that in $b=0.01$ case. Finally the quasi-eigenmode profile which is similar to the eigenfunction in Fig.~\ref{fig:modal}(b) is also formed. The faster growth of perturbation for $b=0.001$ case can be shown more clearly in Fig.~\ref{fig:cyprofnormsb}(b), which plots the original and logarithmic integral intensity versus time. At the end time $t=20/3$, the intensity of perturbation is much greater than that in $b=0.01$ case (about six orders of magnitude compared to Fig.~\ref{fig:cyprofnormlb}(b)), which implies the observable improvement of growth rate induced by the relatively thinner precursor layer. The duration of Stage 1 where the slope is greater is shorter than the $b=0.01$ case, and the curve in Stage 2 is more close to a line with a constant slope, as shown in the logarithmic plot. It indicates the small change of base state in the $b=0.001$ case can make the growth rate keeping unchanged for a longer time.

\begin{figure}
  \includegraphics[width=0.49\textwidth]{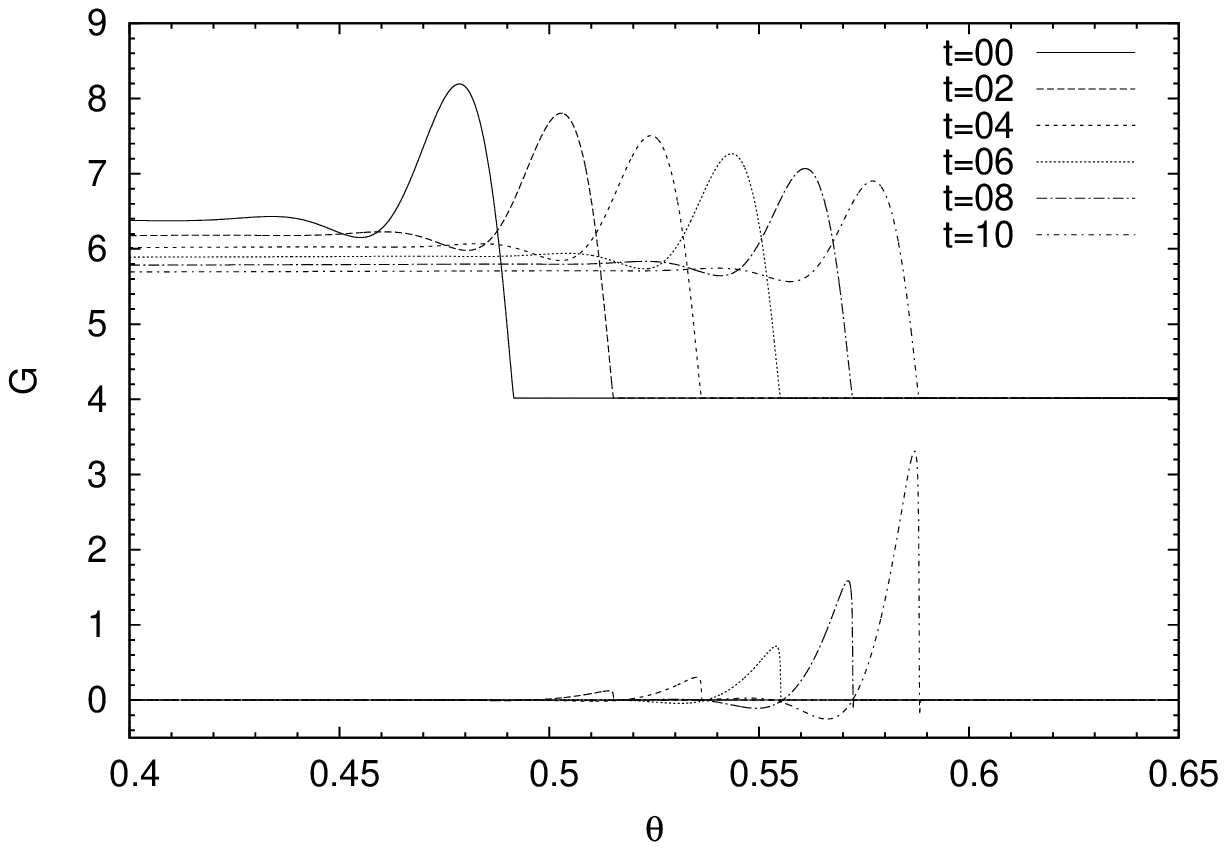}
  \includegraphics[width=0.49\textwidth]{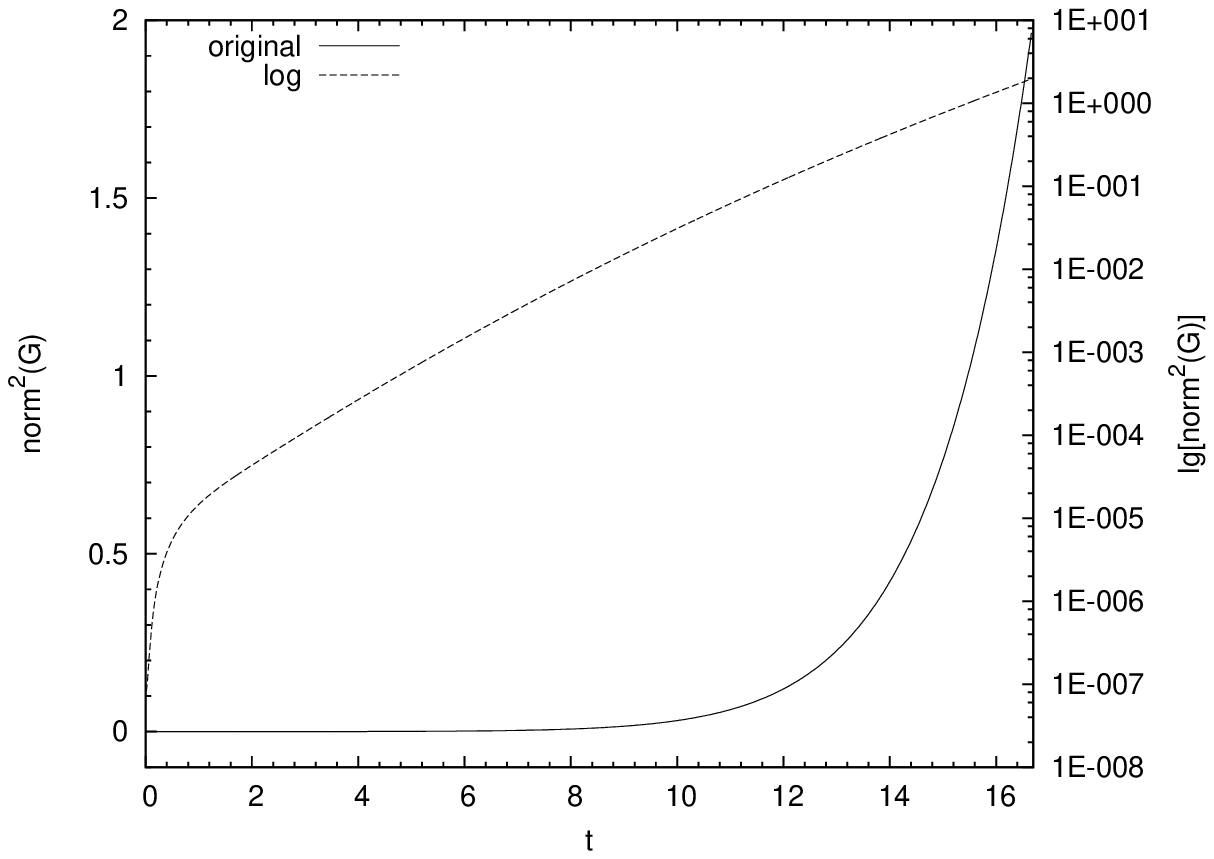}
  \centerline{(a)\hspace{0.45\textwidth}(b)}
  \caption{\label{fig:spprofnorm} (a) Transient profiles of linear perturbation and corresponding axisymmetric capillary wave on spherical surface. The parameters are $b=0.001$, $logBo=6.0$ and $q=40$. The initial perturbation is superimposed at $t=10$. (b) The original and logarithmic integral intensity as a function of time.}
\end{figure}

For spherical problem, the transient profiles of linear perturbation and corresponding axisymmetric capillary wave are shown in Fig.~\ref{fig:spprofnorm}(a). The disturbance time is ranging from $t=0$ to $t=10$. The Bond number is $logBo=6.0$ and the precursor thickness is $b=0.001$. The initial perturbation is superimposed at $t=10$, timing from initial releasing moment. The front location is at $\theta_\uF\approx$0.5-0.6 and the wave number is set to the most unstable value $q=40$. Because the linear perturbation equation (\ref{eq:sppertq}) degenerates into an identical eigenvalue equation, compared to the cylindrical problem, there are inherent similarities for perturbation profiles. At early time, the initial Gaussian perturbation is changed into a quasi-eigenmode. Then this quasi-eigenmode is amplified exponentially. The only difference is that the traveling speed of the quasi-eigenmode is much slower than that in cylindrical problem. Figure \ref{fig:spprofnorm}(b) shows the original and logarithmic plots for integral intensity versus time on spherical surface. Similar to the cylindrical problem, there are also two stages in which Stage 1 represents the nonnormal process related to the initial Gaussian profile, and Stage 2 represents the quasi-eigenmode dominant growth process. But the growth rate in Stage 2 decreases with time clearly and is much smaller than that in cylindrical problem. The former can be also attributed to the explicit change of the base state profiles.

\begin{figure}
  \includegraphics[width=0.49\textwidth]{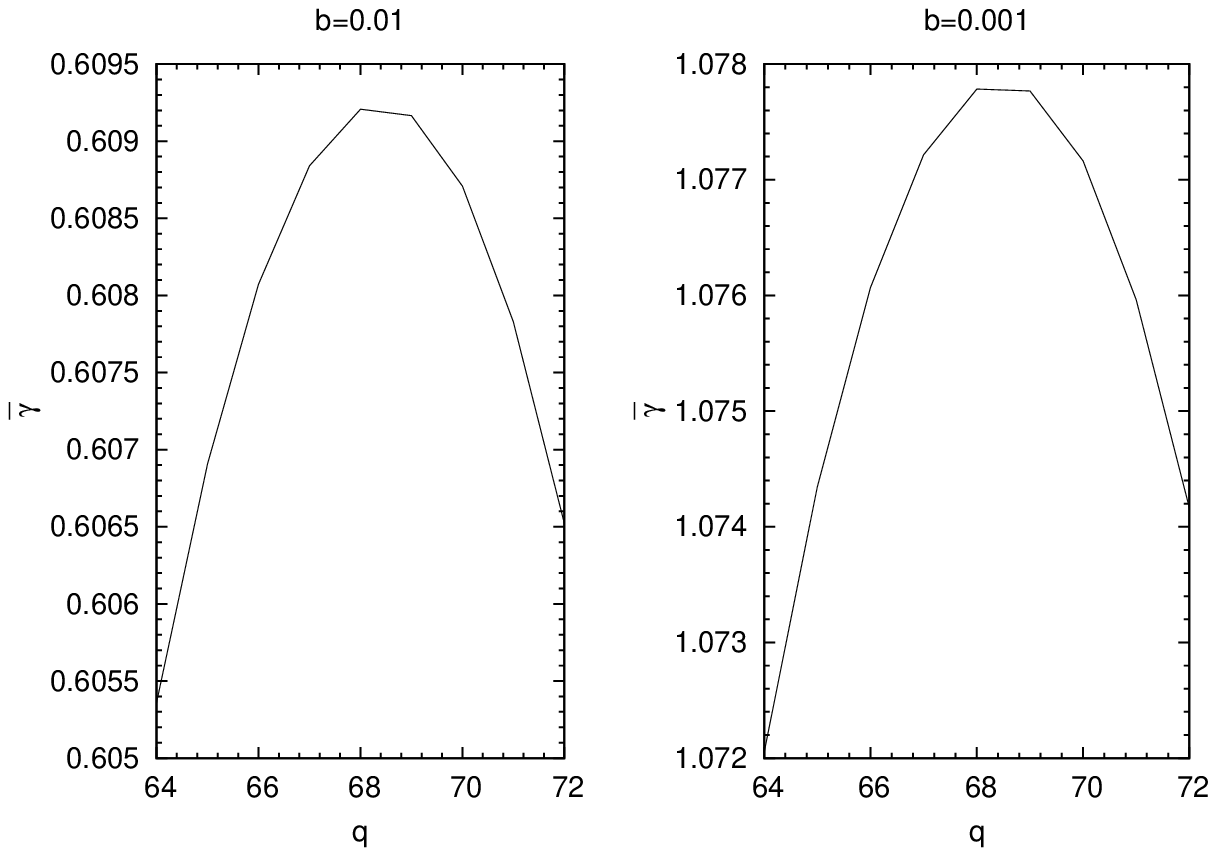}
  \includegraphics[width=0.49\textwidth]{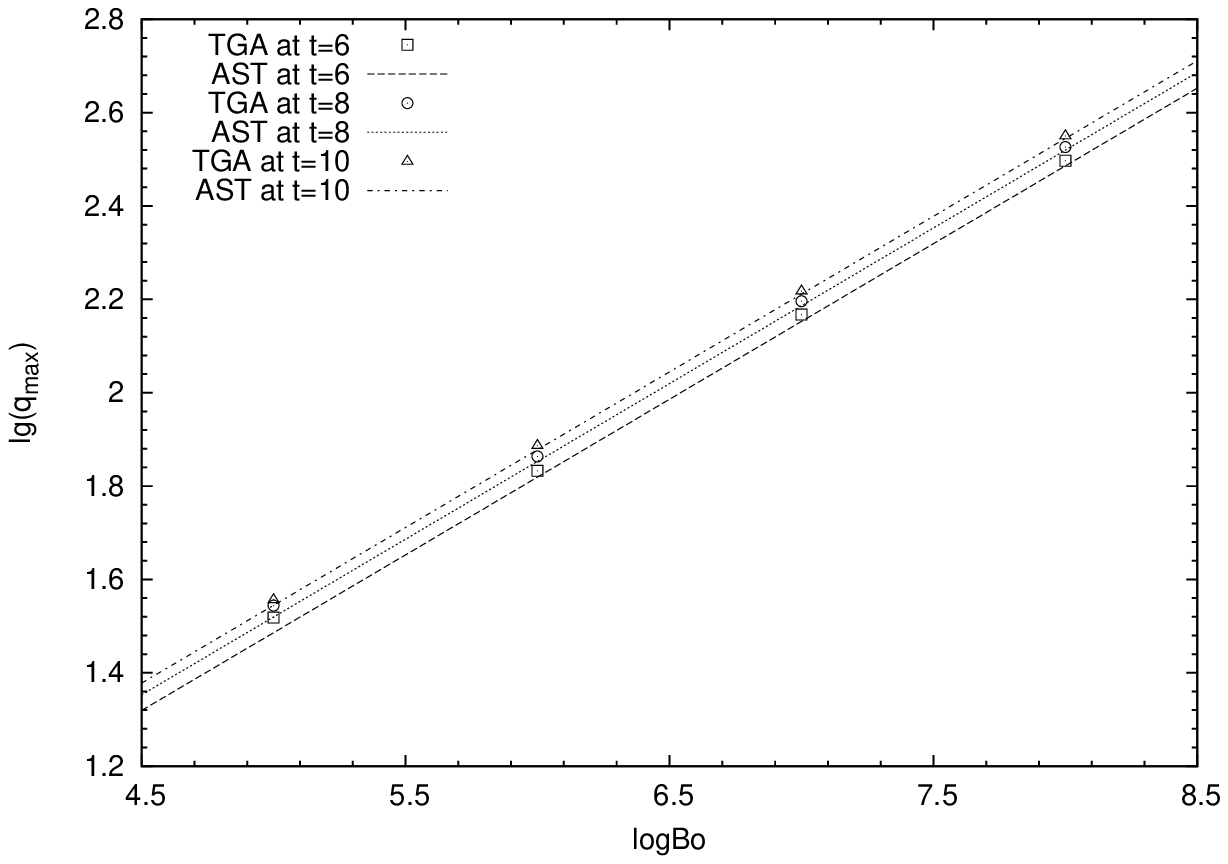}
  \centerline{(a)\hspace{0.45\textwidth}(b)}
  \centerline{\includegraphics[width=0.5\textwidth]{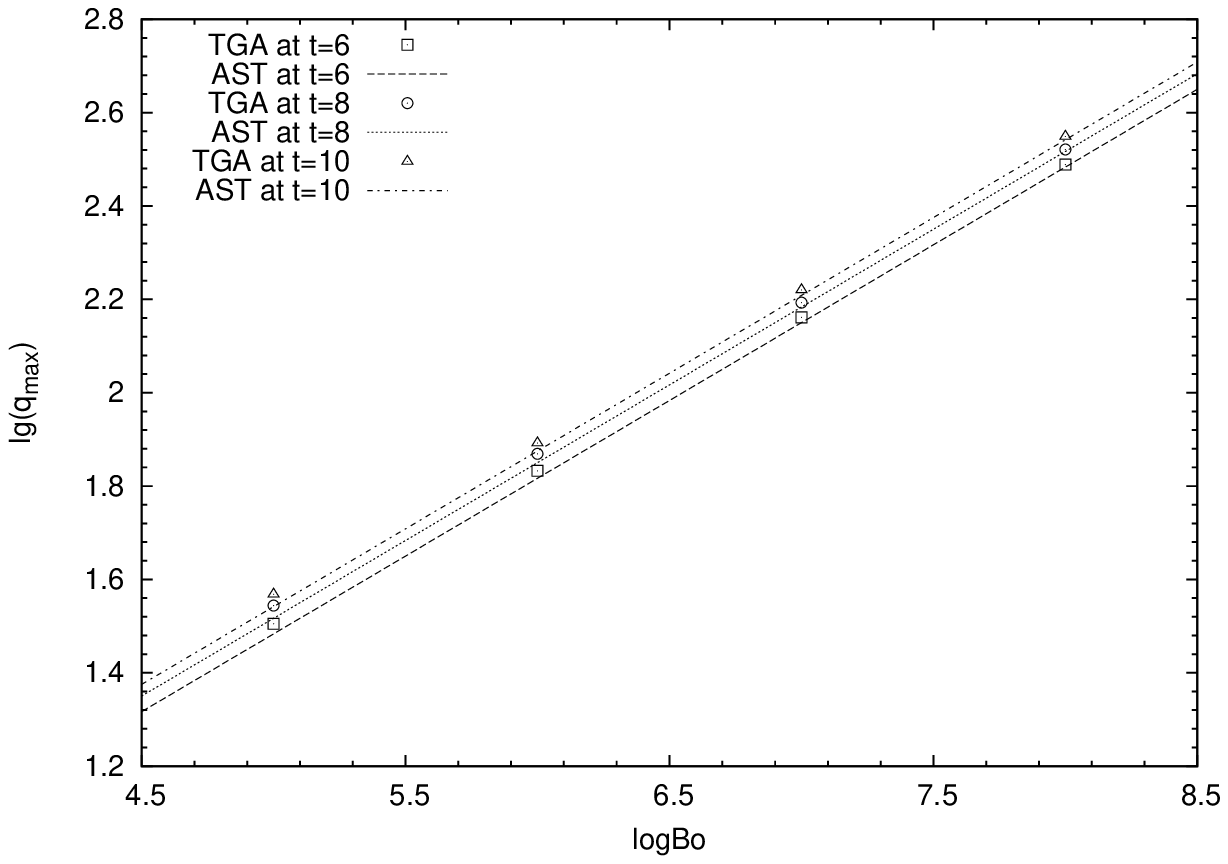}}
  \centerline{(c)}
  \caption{\label{fig:cydisper} (a) The dispersion curve near the most unstable wave number for two precursor thickness $b=0.01$ and $b=0.001$ on cylindrical surface. (b) Log-log plot of the most unstable wave number versus Bond number calculated by TGA as shown via discrete data points. The three base state fronts correspond to three capillary waves on cylindrical surface at $t=6$, $t=8$ and $t=10$, and the precursor thickness is $b=0.01$. The log-log plots calculated from AST are shown as three straight lines. (c) The log-log plots of the most unstable wave number versus Bond number calculated from TGA and AST for a different precursor thickness $b=0.001$.}
\end{figure}

The period $T$ in (\ref{eq:avrate}) should be determined to calculate the average growth rate $\bar{\gamma}$ for constructing a dispersion relationship which is similar to the modal analysis. Because for a given wave number we want to obtain a growth rate which is unrelated to the initial condition, the period should be much greater than the duration in Stage 1 for eliminating the effect of the early-time nonnormal process on the total growth rate and make sure that the dominant quasi-eigenmode has enough time to govern the linear perturbation evolution. Thus the period $T$ is set to ten times of the duration in Stage 1, $T=10T_{\mathrm{stage}1}$, and this condition is not changed in the following computational cases. For cylindrical problem, Fig.~\ref{fig:cydisper}(a) shows the dispersion curves $\bar{\gamma}(q)$ near the most unstable wave number for two precursor thickness $b=0.01$ and $b=0.001$. The value of time and $Bo$ are identical to that in Fig.~\ref{fig:cyprofnormlb}(a) and Fig.~\ref{fig:cyprofnormsb}(a). The average growth rate in the plot is calculated using (\ref{eq:avrate}), and the step size for the wave number is 1.0. The most unstable wave number can be extracted from the peak of the dispersion curve, they are $q=68$ for both $b=0.01$ and $b=0.001$ cases. The relatively thinner precursor layer does not change the most unstable wave number, keeping the unit place, which accords with the prediction of modal analysis, as shown in Fig.~\ref{fig:modal}(a). Note that the most unstable wave number may blue-shift if the range of disturbance time exceeds the limit of $T$. It is not a surprising result because the long time evolution makes the capillary wave much different from the base state at the moment when the initial perturbation is imposed. The front location is greater and front thickness is thinner, compared to the original front, which introduce a greater most unstable wave number. But in practice, the duration of linear perturbation evolution is short and the evolution may quickly access the nonlinear stage. The selection of period $T$ can be considered as a terminal criterion for a typical linear stage.

To verify the accuracy of formula (\ref{eq:cymodalqmax}) derived from the asymptotic theory and modal analysis, three fronts corresponding to three capillary waves at $t=6$, $t=8$ and $t=10$ are selected as the base states, and the precursor thickness is $b=0.01$. The value of $Bo$ is varied to examine the effects of Bond number on the most unstable wave number. The data points concerning the most unstable wave number versus $Bo$ are shown as a log-log plot in Fig.~\ref{fig:cydisper}(b). The range of $Bo$ is from $logBo=5.0$ to $logBo=8.0$ with the step size 1.0. The three lines displayed in Fig.~\ref{fig:cydisper}(b) is the plots of $\lg(q_\umax)$ versus $logBo$ computed using (\ref{eq:cymodalqmax}) according to asymptotic theory (AST), the front location $\theta_\uF$ and front thickness $h_\uF$ as the coefficients in (\ref{eq:cymodalqmax}) are extracted from the outer solution of the corresponding capillary wave. The slopes of the lines are well fitting the data points. The values of data points show quantitative consistency with the power law $Bo^{1/3}$ derived from modal analysis. According to (\ref{eq:cymodalqmax}), the intercept of the lines is $\lg(\frac{\sin^{1/3}\theta_\uF}{2h_\uF^{1/3}})$. The values of $q_\umax$ at a given $Bo$ calculated from TGA decrease with time, as shown from the data points. It indicates $q_\umax$ increases as the front location $\theta_\uF$ increases, which agree with the intercept of lines. There are more differences between the data points and lines when $logBo=5.0$, these discrepancies are not obvious when $logBo$ is greater than 6.0. It indicates the power law (\ref{eq:cymodalqmax}) is more valid in high $Bo$ case, corresponding to an asymptotic behavior in high $Bo$ limit. Figure \ref{fig:cydisper}(c) shows the log-log plot of the most unstable wave number versus $Bo$ for the precursor thickness $b=0.001$, the other parameters are identical to the $b=0.01$ case. A good fitting relationship between the data points from TGA and the lines from AST is also illustrated for the $b=0.001$ case. The data points in high $Bo$ range is more close to the lines compared to the low $Bo$ case, which also implies an asymptotic behavior as $Bo$ increases.

\begin{figure}
  \includegraphics[width=0.49\textwidth]{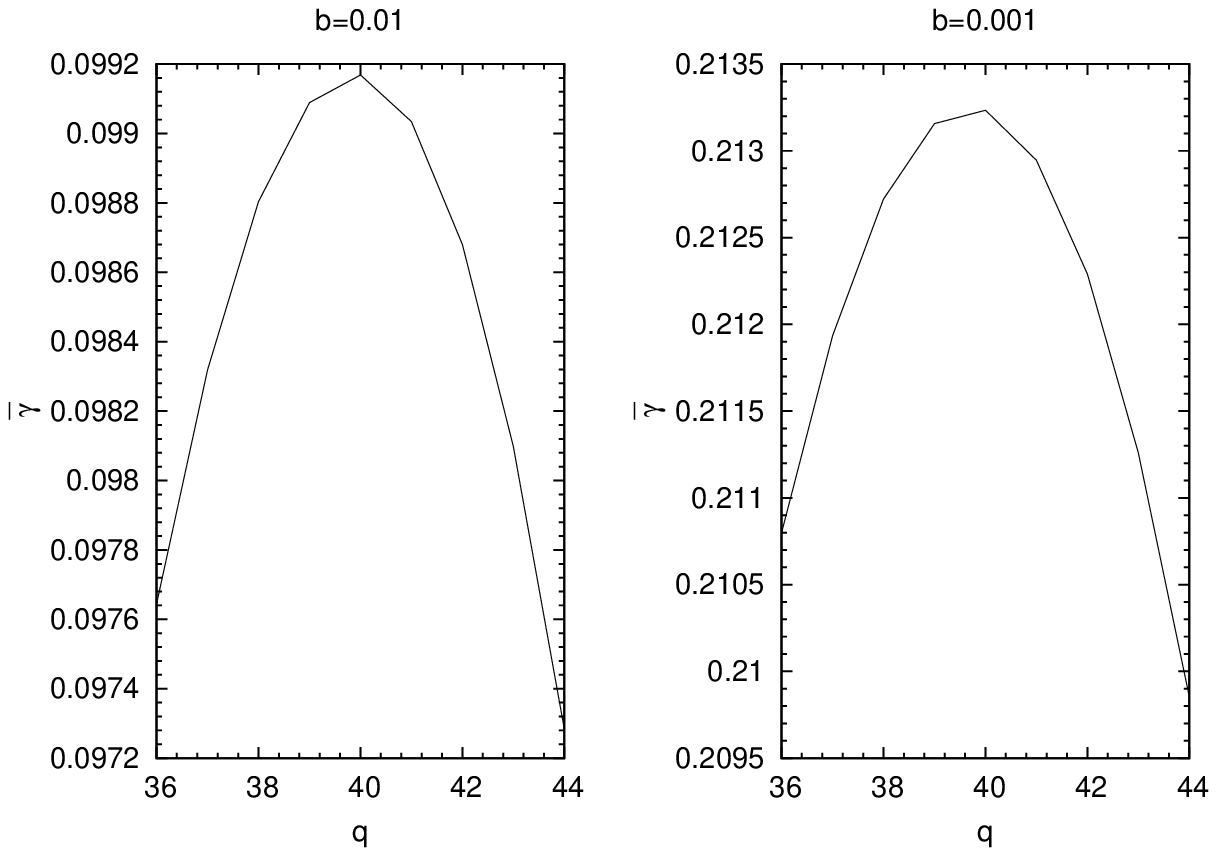}
  \includegraphics[width=0.49\textwidth]{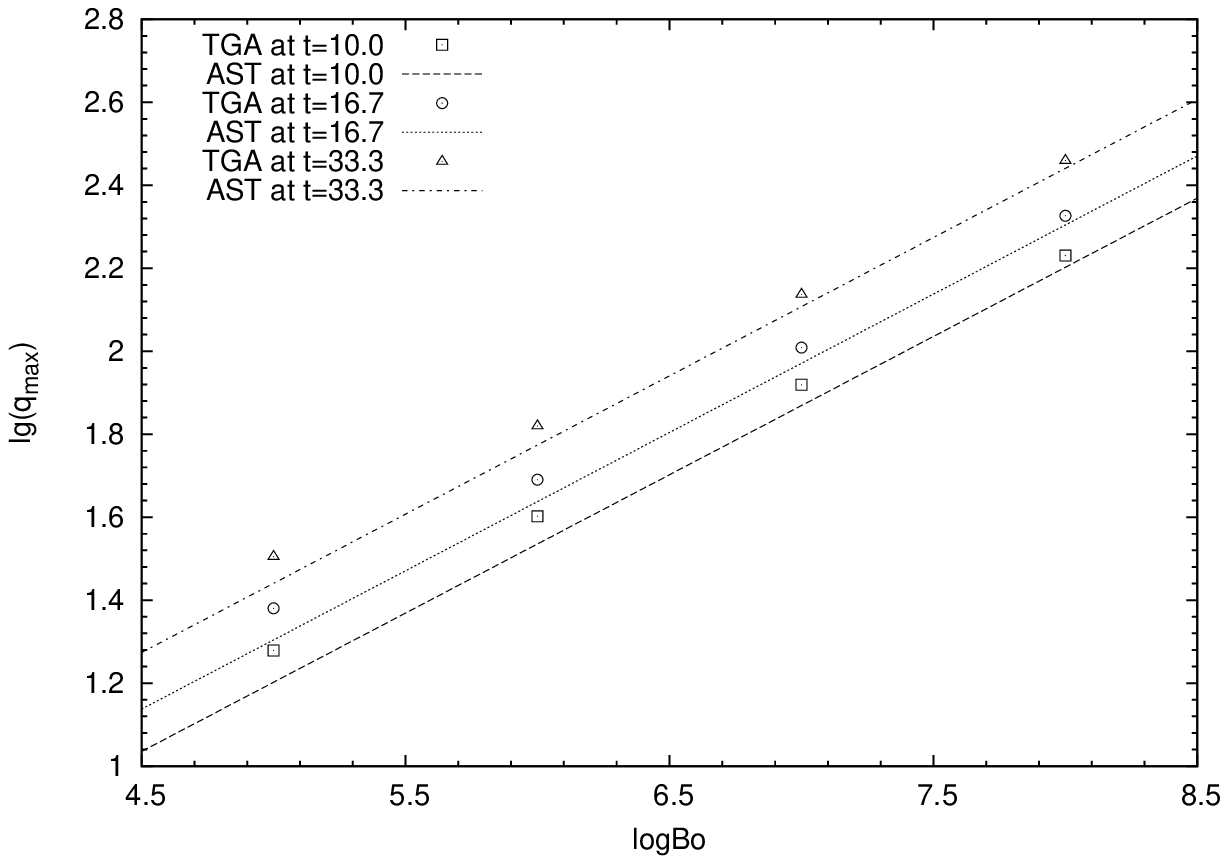}
  \centerline{(a)\hspace{0.45\textwidth}(b)}
  \centerline{\includegraphics[width=0.5\textwidth]{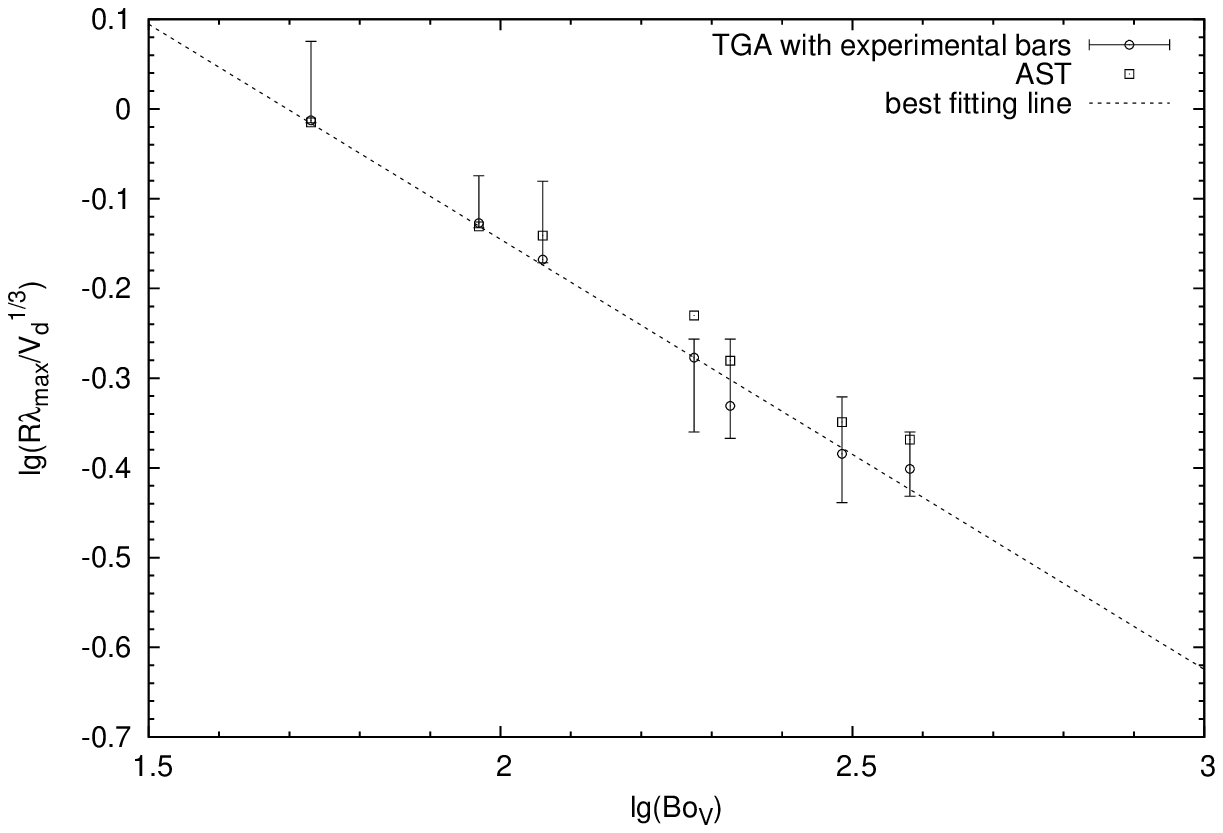}}
  \centerline{(c)}
  \caption{\label{fig:spdisper} (a) The dispersion curve near the most unstable wave number for two precursor thickness $b=0.01$ and $b=0.001$ on spherical surface. (b) The log-log plots of the most unstable wave number versus Bond number for spherical problem, where the discrete data points represent the numerical results from TGA and the straight lines represent the analytical results from AST. The three base-state profiles are at $t=10$, $t=50/3$ and $t=100/3$, and the precursor thickness is $b=0.001$. (c) The most unstable wavelength versus volumetric Bond number on logarithmic scales for spherical problem, both the results calculated from TGA (circle) and AST (square) are shown as discrete data points, the minimum and maximum wavelength in the repeated tests for each case are plotted using the error bars. The best fitting line for the TGA results is shown as a straight line.}
\end{figure}

Figure \ref{fig:spdisper}(a) shows the dispersion curve $\bar{\gamma}(q)$ near the most unstable wave number on spherical surface. For the left and right plot, the precursor thickness is $b=0.01$ and $b=0.001$, respectively. The step size for the wave number is 1 (note that the wave number is integer for spherical problem) for both the two plots. The Bond number is $logBo=6.0$ and the initial perturbation is superimposed on an axisymmetric capillary wave at $t=10$ , corresponding to the parameters in Fig.~\ref{fig:spprofnorm}(a). The most unstable wave number is $q=40$ for both $b=0.01$ and $b=0.001$ cases. Similar to the cylindrical problem, the most unstable wave number is insensitive to the change of precursor thickness, but the decrease of precursor thickness can improve the average growth rate observably. Figure \ref{fig:spdisper}(b) shows the log-log plots of the most unstable wave number versus $Bo$ for spherical problem, in which the discrete data points represent the results from TGA and the lines represent theoretical relationship computed using formula (\ref{eq:spmodalqmax}). The three base-state profiles are at $t=10$, $t=50/3$ and $t=100/3$, and the precursor thickness is set to a relatively thinner value $b=0.001$. The range of $Bo$ is from $logBo=5.0$ to $logBo=8.0$, same as the cylindrical problem. The values of data points agree well with the lines from AST when $logBo=8.0$, which verifies the asymptotic behavior at high $Bo$. But compared to the cylindrical problem, the differences between TGA and AST become more significant as $Bo$ decreases. The same differences of the capillary waves between the composite solutions and direct solutions are also discovered on spherical surface at relatively low $Bo$.\citep{Hou2017} These findings can be attributed to a same mechanism: compared to the cylindrical problem, there are more first order $O(Bo^{-1/3})$  terms in the inner form of the linear perturbation equation for spherical problem (see (\ref{eq:sppertqin})), which makes the leading order asymptotic theory less accurate.

\subsection{Experimental evidence}

There are a series of experiments reported by \citet{Takagi2010} which record the fingering number in typical coating flows on upper sphere. The fluid is golden syrup with the physical properties $\rho=1400\mathrm{kg/m^3}$, $\nu=0.045\mathrm{m^2/s}$ and $\sigma=0.078\mathrm{N/m}$. Constant volume fluid drips down a vinyl beach ball of radius $R=24\mathrm{cm}$, and the uniform initial film is located at a top region of radius 5cm on the sphere. The authors defined a volumetric Bond number $Bo_{\mathrm{V}}$ to study the instability feature depending on released fluid volume
\begin{equation*}
Bo_{\mathrm{V}}=\rho gV_{\ud}/\sigma R
\end{equation*}
where $V_\ud$ is the dimensional fluid volume. Different released volume represents different thickness of initial film, i.e., different thickness scale ($H$ in (8)). The relationship between $Bo$ defined in (\ref{eq:bo}) and $Bo_{\mathrm{V}}$ in \citet{Takagi2010} is
\begin{equation}\label{eq:bovbo}
Bo=\rho gR^{3}/\sigma H=\rho^{2}g^{2}R^{2}S/(\sigma^{2}Bo_{\mathrm{V}})
\end{equation}
where $S$ is the area of the region where the film is located at initial moment. The experiments were conducted over a range of $Bo_{\mathrm{V}}$ from 1.7 to 2.6 on logarithmic scales, corresponding to $logBo\approx4.6-5.4$ defined in present paper. The time when the initial disturbances are imposed on the axisymmetric capillary wave is not known in the experimental environment, but the authors record the flow lengths $\theta_{\mathrm{N}}$ when the front begins to rupture for each experimental case. We capture the mean values of the flow lengths as a function of $Bo_{\mathrm{V}}$ as shown in Table \ref{tab:thetan}. This front location can be used to determine the onset of fingering instability. The initial Gaussian perturbation is superimposed onto the base state immediately when the front location $\theta_\uF$ arrives at $\theta_{\mathrm{N}}$. The most unstable wavelength against $Bo_{\mathrm{V}}$ are given by a figure shown in \citet{Takagi2010}. To compare with the experimental data conveniently, a dimensionless most unstable wavelength is defined as
\begin{equation}\label{eq:spwlmax}
\lambda_{\umax}=\frac{2\pi\sin\theta_{\uF}}{q_{\umax}}
\end{equation}
Figure \ref{fig:spdisper}(c) shows the most unstable wavelength versus $Bo_{\mathrm{V}}$ on logarithmic scales, both the results calculated from TGA and AST are shown as discrete data points, the minimum and maximum most unstable wavelength for each case observed in the repeated tests are plotted using the error bars. The value of most unstable wavelength can not be fixed in repeated independent trials. It's mainly due to the randomness of the time when the initial perturbation is imposed (onset of the fingering instability). This randomness can be proved via another plot of $\theta_{\mathrm{N}}$ against $Bo_{\mathrm{V}}$ shown in \citet{Takagi2010}, because the flow length $\theta_{\mathrm{N}}$ is not fixed in repeated experiments too (we choose an average value as the flow length in Table \ref{tab:thetan}). Note that the scale of the wavelength ($V_\ud^{1/3}$) in \citet{Takagi2010} is different from the length scale (i.e., $R$) used here, the value of $\lambda_\umax$ computed using (\ref{eq:spwlmax}) should be multiplied a factor $R/V_\ud^{1/3}$ as the ordinate in the plot. The line best fitting the TGA data points scales like $Bo_{\mathrm{V}}^{-0.47}$, as shown in Fig.~\ref{fig:spdisper}(c), the exponent -0.47 is slightly greater than -0.52 which is given in \citet{Takagi2010}. The formula of dimensionless most unstable wavelength derived from AST can be expressed as
\begin{equation}\label{eq:spmodalwlmax}
\lambda_{\umax}=\frac{4\pi h_{\uF}^{1/3}}{\sin^{1/3}\theta_{\uF}Bo^{1/3}}
\end{equation}
The differences between the data points from AST and that from TGA can be attributed to the leading order accuracy of the asymptotic theory used here. Apparently the range of $logBo$ (4.6-5.4) in the experimental cases is not sufficiently large, but most data points from AST fall into the error bars discovered in the experiments, and the asymptotic theory and modal analysis still provide reasonable most unstable wavelength compared to the TGA and experimental results. It indicates that the reported experimental cases satisfy the asymptotic conditions in the sense of leading order.

\begin{table}
\caption{\label{tab:thetan}Records of the mean flow lengths at the onset of the fingering instability for different volumetric Bond number}
\begin{ruledtabular}
  \begin{tabular}{cc}
  $\lg(Bo_\mathrm{V})$ & $\theta_{\mathrm{N}}$\\
  \hline
  1.731 & 0.416\\
  1.970 & 0.550\\
  2.060 & 0.564\\
  2.275 & 0.702\\
  2.326 & 0.799\\
  2.485 & 0.958\\
  2.582 & 1.011\\
  \end{tabular}
\end{ruledtabular}
\end{table}

\subsection{The partial wetting cases}

\begin{figure}
  \includegraphics[width=0.49\textwidth]{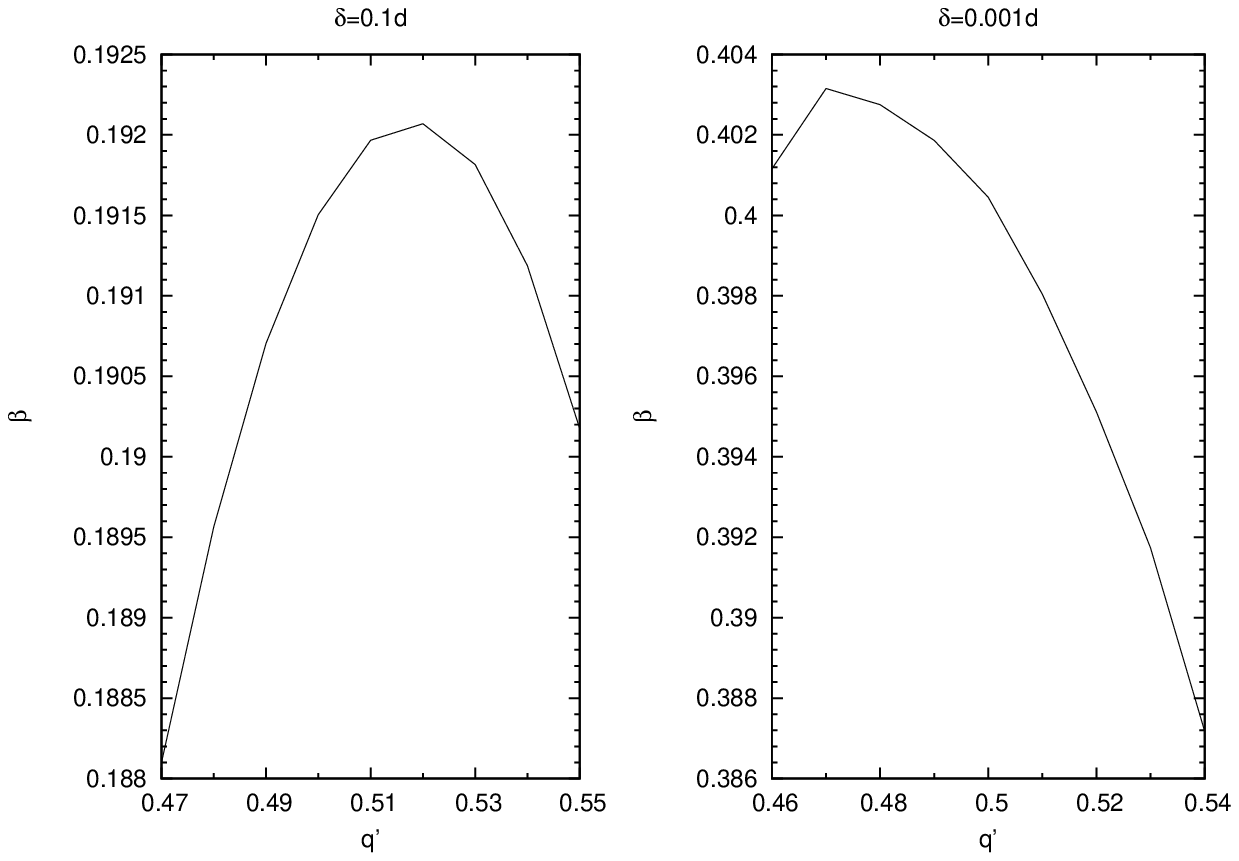}
  \includegraphics[width=0.49\textwidth]{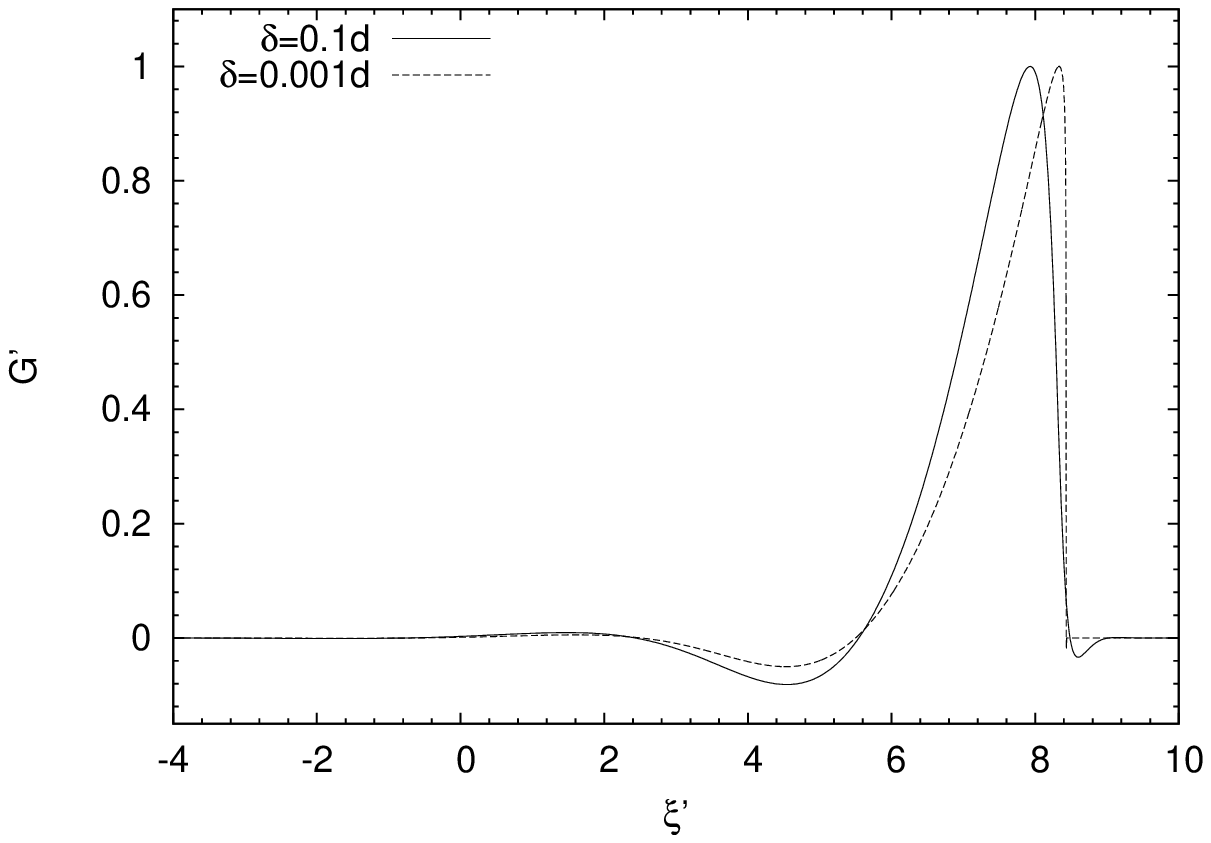}
  \centerline{(a)\hspace{0.45\textwidth}(b)}
  \caption{\label{fig:modaldjp} (a) The dispersion curve near the most unstable region for two precursor parameters $\delta=0.1$ and $\delta=0.001$ in partial wetting cases, the dimensionless contact angle parameter $K$ is 10 for $\delta=0.1$ case and 1000 for $\delta=0.001$ case, the suffix `d' in the legend denotes the disjoining pressure is taken into calculation. (b) The comparison of eigenfunction profiles corresponding to most unstable inner wave number between the $\delta=0.1$ case and $\delta=0.001$ case in partial wetting cases.}
\end{figure}

In this subsection, we will focus on the partial wetting cases in which the disjoining pressure is operative. For modal analysis, the eigenvalue equation (\ref{eq:eigen}) is replaced by (\ref{eq:eigendjp}) to take the disjoining pressure terms into account. Figure~\ref{fig:modaldjp}(a) shows the dispersion curve $\beta(q')$ near the most unstable region for two relative precursor thickness $\delta=0.1$ and $\delta=0.001$. The values of $Ca$, $n$, $m$ and $\theta_\ue$ in (\ref{eq:djpcoin}) are fixed, and the dimensionless contact angle parameter $K$ only varies with $\delta$ and is set to 10 and 1000, corresponding to $\delta=0.1$ case and $\delta=0.001$ case, respectively. The most unstable wave number appears at $q'=0.52$ for $\delta=0.1$ case, blue-shifts only 0.03 compared to the value solved from (\ref{eq:eigen}) as shown in Fig.~\ref{fig:modal}(a). As the precursor thickness decreases, the most unstable wave number appears at $q'=0.47$ for $\delta=0.001$ case, which is identical to that in complete wetting cases. But keeping the accuracy to the first decimal place, the value of most unstable wave number can be still taken as $q'_\umax\approx0.5$ for each case. It indicates the disjoining pressure may slightly increase $q'_\umax$ in case of relatively thicker precursor layer , but the value of $q'_\umax$ is insensitive to the addition of disjoining pressure if the precursor layer is sufficiently thin. Actually the relative precursor thickness is thin (in microscopic scale) in most partial wetting cases. The profiles of eigenfunction $G'$ corresponding to most unstable wave number are shown in Fig.~\ref{fig:modaldjp}(b), for both the $\delta=0.1$ and $\delta=0.001$ cases. Note that the secondary minimum near the apparent contact line for both profiles is weakened, which is one effect of the disjoining pressure. The peak location of eigenfunction for $\delta=0.001$ case moves right compared to $\delta=0.1$ case, which is different from that displayed in Fig.~\ref{fig:modal}(b). It can be attributed to another effect in partial wetting cases.

\begin{figure}
  \includegraphics[width=0.49\textwidth]{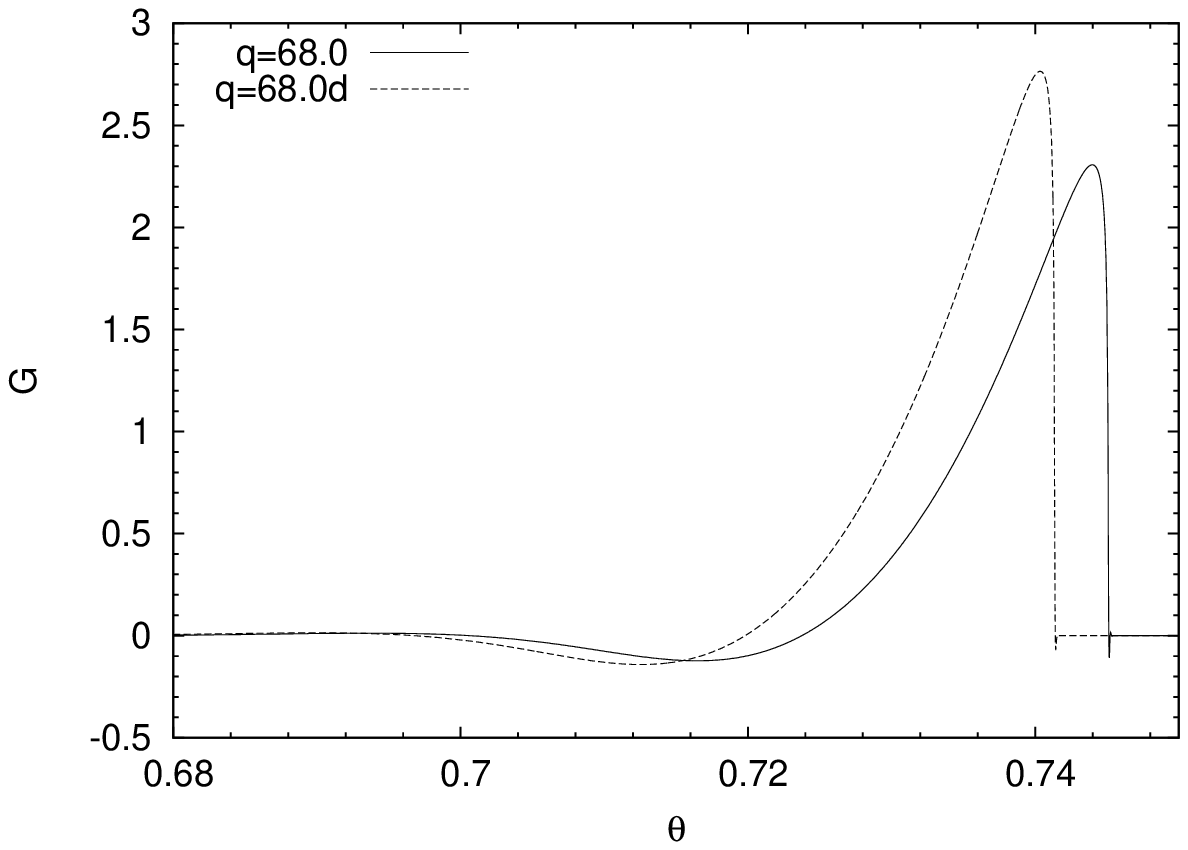}
  \includegraphics[width=0.49\textwidth]{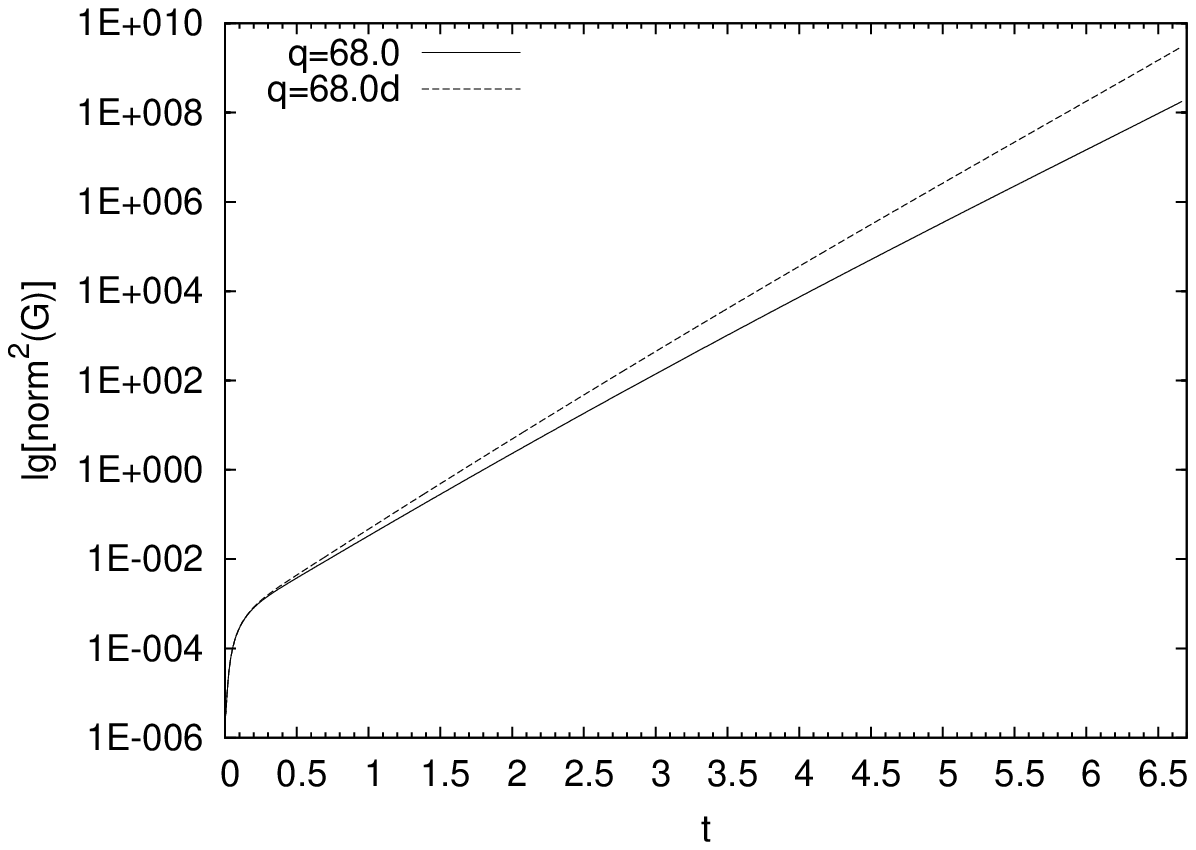}
  \centerline{(a)\hspace{0.45\textwidth}(b)}
  \caption{\label{fig:cyprofnormdjp} (a) The linear perturbation profiles solved from the cylindrical perturbation equations in complete and partial wetting cases. The parameters of the base state are identical to that in Fig.~\ref{fig:cyprofnormsb}(a). The perturbation profiles are captured at $t=1$ after the initial perturbation is imposed, and the wave number is $q=68$. The parameters for disjoining pressure are $(n,m)=(3,2)$, $\theta_\ue=0.5$, and $\epsilon=0.015$. (b) The logarithmic integral intensity versus time for the wave number $q=68$ in complete and partial wetting cases.}
\end{figure}

\begin{figure}
  \centerline{\includegraphics[width=0.6\textwidth]{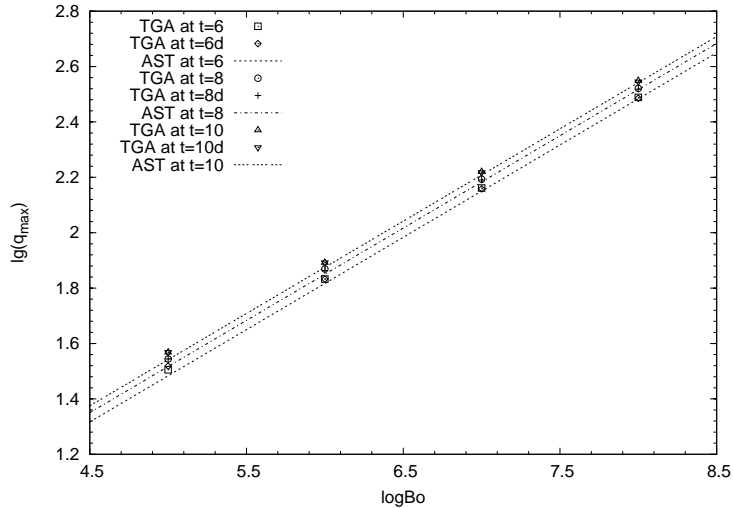}}
  \caption{\label{fig:cyasymdjp} The effect of disjoining pressure on the asymptotic behavior for cylindrical problem. The discrete data points are TGA results in complete and partial wetting cases. The lines are calculated using AST. The base-state fronts are at $t=6$, $t=8$ and $t=10$ with precursor thickness $b=0.001$ and the parameters for disjoining pressure are same as that in Fig.~\ref{fig:cyprofnormdjp}(a).}
\end{figure}

Back to the transient growth analysis, the linear perturbation equations (\ref{eq:cypertqdjp}) and (\ref{eq:sppertqdjp}) which include the disjoining pressure terms can be used as the canonical equations. Figure \ref{fig:cyprofnormdjp}(a) shows the comparison of the perturbation profiles solved from (\ref{eq:cypertq}) and (\ref{eq:cypertqdjp}). The parameters are identical to that in Fig.~\ref{fig:cyprofnormsb}(a) ($logBo=6.0$ and $b=0.001$). The base state corresponds to the two-dimensional capillary wave at $t=6$, the perturbation profiles are captured at $t=1$ after the initial perturbation is imposed. The parameters in equation (\ref{eq:djpd}) are $(n,m)=(3,2)$, $\theta_\ue=0.5$, and $\epsilon=0.015$. The most unstable wave number which is calculated in partial wetting cases does not shift compared to the case in Fig.~\ref{fig:cyprofnormsb}(a) ($q_{\umax}$ is also 68). The perturbation amplitude in $q=68\ud$ case are obviously greater than that in $q=68$ case, as demonstrated in Fig.~\ref{fig:cyprofnormdjp}(a). This feature can be seen more clearly in Fig.~\ref{fig:cyprofnormdjp}(b), which shows the logarithmic plots of the perturbation profiles' integral intensity versus time. The intensity of $q=68\ud$ profile is much greater than intensity of $q=68$ profile at the end time $t=20/3$. It implies that the disjoining pressure can increase the linear growth rate observably. Figure \ref{fig:cyasymdjp} shows the log-log plots of the most unstable wave number versus $Bo$. In partial wetting cases, the three base state fronts are selected corresponding to the two-dimensional capillary waves solved from (\ref{eq:tddjpeq}). The values of time and precursor thickness are identical to that in Fig.~\ref{fig:cydisper}(c). The discrete data points are TGA results in complete and partial wetting cases. The lines are calculated using AST (see (\ref{eq:cymodalqmax})) which is independent on the wetting model. Compared to the original TGA results in complete wetting cases, it is clear that the most unstable wave number is insensitive to the additional disjoining pressure terms. The TGA results calculated in partial wetting cases also approach the lines asymptotically as $Bo$ increases. The asymptotic behavior with the increase of $Bo$ is not affected by the wetting process.


\begin{figure}
  \includegraphics[width=0.49\textwidth]{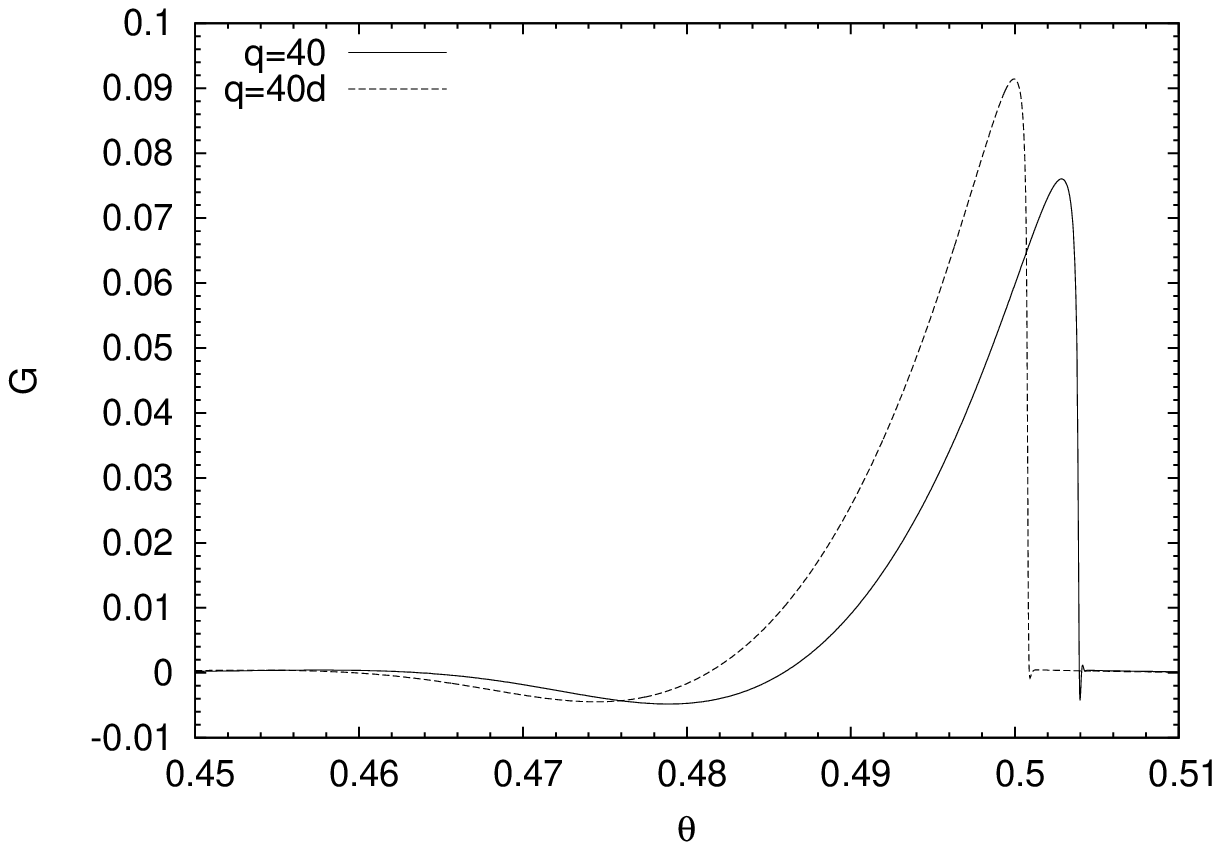}
  \includegraphics[width=0.49\textwidth]{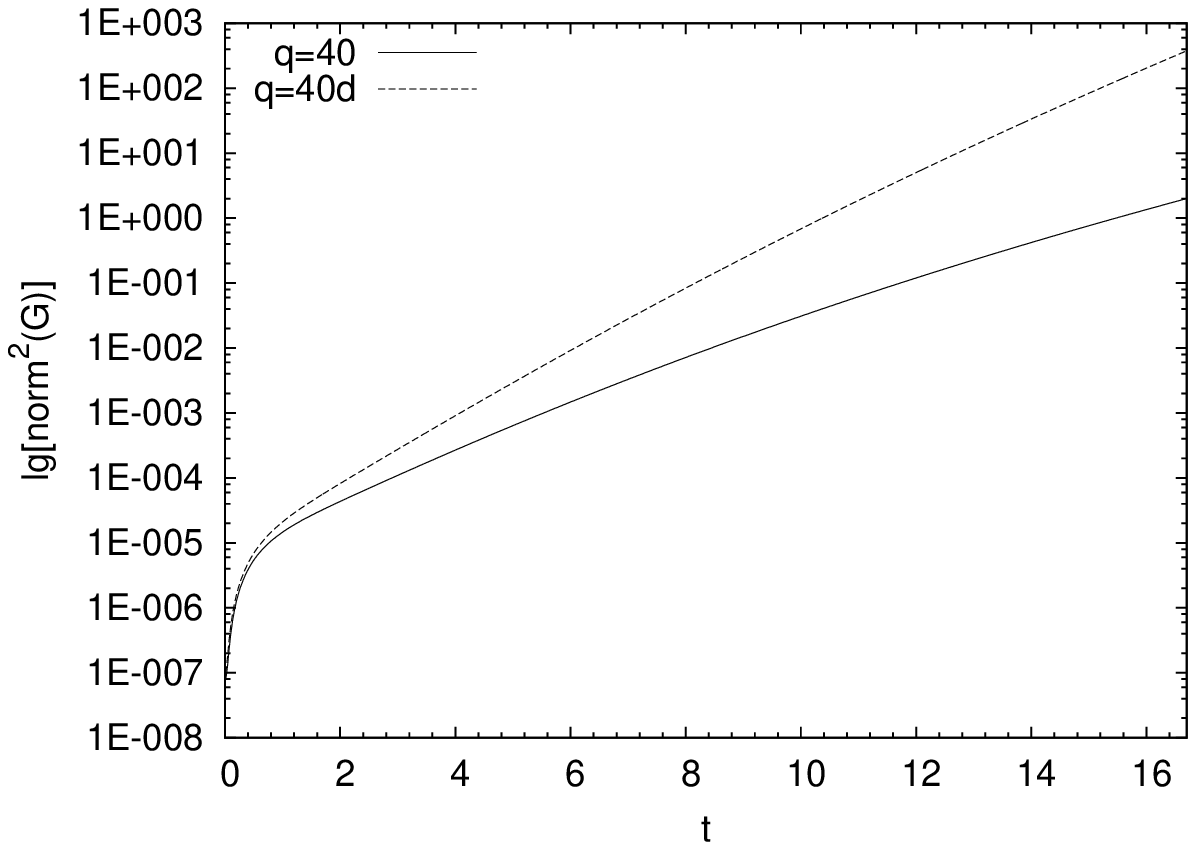}
  \centerline{(a)\hspace{0.45\textwidth}(b)}
  \caption{\label{fig:spprofnormdjp} (a) The linear perturbation profiles solved from the spherical perturbation equations in complete and partial wetting cases. The parameters of the base state are identical to that in Fig.~\ref{fig:spprofnorm}(a). The perturbation profiles are captured at $t=1$ after the initial perturbation is imposed, and the wave number is $q=40$. The parameters for disjoining pressure are $(n,m)=(3,2)$, $\theta_\ue=0.5$, and $\epsilon=0.015$. (b) The logarithmic integral intensity versus time for the wave number $q=40$ in complete and partial wetting cases.}
\end{figure}

\begin{figure}
  \centerline{\includegraphics[width=0.6\textwidth]{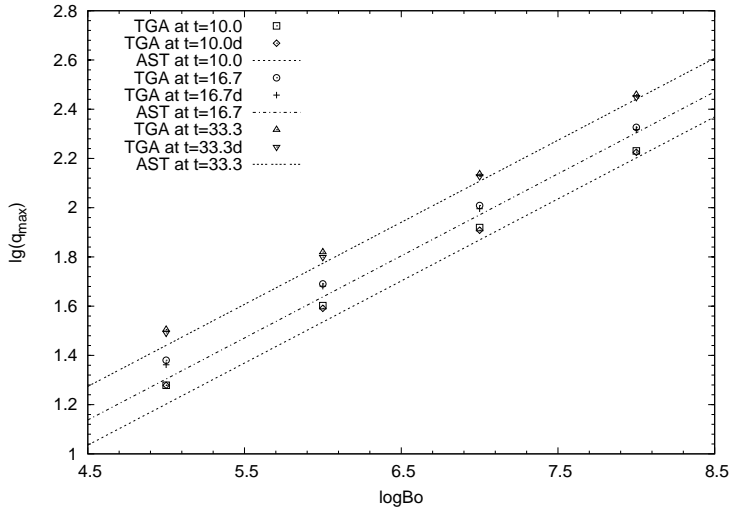}}
  \caption{\label{fig:spasymdjp} The effect of disjoining pressure on the asymptotic behavior for spherical problem. The discrete data points are TGA results in complete and partial wetting cases. The lines are calculated using AST. The base-state fronts are at $t=10$, $t=50/3$ and $t=100/3$ with precursor thickness $b=0.001$ and the parameters for disjoining pressure are same as that in Fig.~\ref{fig:spprofnormdjp}(a).}
\end{figure}

For spherical problem, Fig.~\ref{fig:spprofnormdjp}(a) shows the comparison of the perturbation profiles solved from (\ref{eq:sppertq}) and (\ref{eq:sppertqdjp}). The parameters of the base state are identical to that in Fig.~\ref{fig:spprofnorm}(a) ($logBo=6.0$, $b=0.001$ and $t=10$). The parameters of the disjoining pressure are identical to the cylindrical problem ($(n,m)=(3,2)$, $\theta_\ue=0.5$, and $\epsilon=0.015$). The two perturbation profiles are captured at $t=1$ after the initial perturbation is imposed. A wave number $q=40$ which is most unstable for both two cases is selected. The same improvement of perturbation amplitude induced by disjoining pressure can be observed, compared to the cylindrical problem. The corresponding growth rate for $q=40\ud$ case is definitely greater than that in $q=40$ case, which can be seen more clearly in the log-log plots of integral intensity versus time as shown in Fig.~\ref{fig:spprofnormdjp}(b). Figure \ref{fig:spasymdjp} shows the comparison of asymptotic behavior between the complete and partial wetting cases. The base state fronts are selected corresponding to the axisymmetric capillary waves in complete and partial wetting cases, and the time and precursor thickness are identical to that in Fig.~\ref{fig:spdisper}(b). The lines from AST are calculated using (\ref{eq:spmodalqmax}). The most unstable wave number is insensitive to the addition of disjoining pressure, which is analogous to that in cylindrical problem as shown in Fig.~\ref{fig:cyasymdjp}. When $Bo$ increases, the same asymptotic behavior is also observed in partial wetting cases, compared to that in complete wetting cases.


\section{\label{sec:con}Conclusions}

We study the fingering instability in typical gravity driven coating flows on upper cylinder and sphere. These phenomena can be described as the transverse instability of a two-dimensional or axisymmetric capillary wave. Based on the leading order governing equation for coating flow down cylindrical or spherical surface, the two-dimensional or axisymmetric evolution equation and corresponding linear perturbation equation are derived for cylindrical problem and spherical problem, respectively. A common dimensionless Bond number is introduced after the nondimensionalization of the governing equations. A capillary wave which is the solution of the two-dimensional or axisymmetric evolution equation can be used as the base state. Then an initial perturbation which is localized in the capillary ridge is imposed on the base state profile at a certain moment and may induce subsequent fingering. A disjoining pressure model which is compatible with a precursor film condition is added into the capillary wave equation and linear perturbation equation to model the partial wetting process in which a finite contact angle exists.

We focus on the high $Bo$ flow which can be considered as situations that correspond to relatively thinner liquid film flow on larger sized substrate. The high $Bo$ limit of the linear perturbation equations is studied by an asymptotic theory, using the method of matched asymptotic expansions. Because the linear perturbation for a high $Bo$ flow is always localized near the moving front, the form of linear perturbation equation in the inner region is highlighted. It is proved that, for the cylindrical and spherical problems, the linear perturbation equations in the inner region may degenerate into a common form which is identical to the eigenvalue problem on planar surface, if the duration of linear growth is much smaller than the time scale of capillary wave and the wave number is sufficiently large. Two analytical formulae concerning the most unstable wave number on cylindrical and spherical surfaces which differ only in one factor are derived. These formulae can be described as power laws and applied to estimating the fingering number which may evolve in the nonlinear stage, if the values of $Bo$, front location, front thickness and the most unstable inner wave number at a certain moment are given. The asymptotic theory also provides a method to calculate the linear growth rate from the eigenvalue. The method for introducing the partial wetting process into the asymptotic theory is also discussed. The disjoining pressure terms will appear on the right side of the inner modal equation using a common form for both the cylindrical and spherical problems.

Some adaptive methods are proposed for numerically solving the modal equations and the linear perturbation (non-modal) equations. Based on the numerical solutions for modal and non-modal equations, the methods of modal analysis and transient growth analysis can be used to demonstrate the results for high $Bo$ flow. According to the eigenvalues calculated from the modal equation, a dispersion curve can be established to extract the most unstable inner wave number. Then the most unstable outer wave number is calculated using the formulae from the asymptotic theory. The value of most unstable inner wave number is slightly affected by the relative precursor thickness, but the growth rate (eigenvalue) is sensitive to the decrease of the relative precursor thickness. The transient growth can be calculated by solving the linear perturbation equation directly after imposing a Gaussian initial perturbation. A dispersion curve which is similar to the modal analysis can be constructed using a method of average growth rate for a given wave number. Some features concerning the typical profiles of linear perturbation and the evolution of the integral intensity (square 2-norm) are shown for cylindrical and spherical problem, respectively. Since the linear perturbation equations degenerate into an identical eigenvalue equation, there are inherent similarities for perturbation profiles on cylindrical and spherical surfaces. It is found that there are two stages in the linear perturbation evolution, a non-modal stage which is related to the initial condition at early time and a quasi-eigenmode dominated stage which is similar to the eigenfunction in modal analysis at later time. The growth rate in the quasi-eigenmode dominated stage may not be a constant because the time evolution of the capillary wave profile cannot be neglected if the linear evolution sustains for a long time. Compared to the modal analysis, the effect of precursor thickness obtained from TGA is identical (insensitive most unstable wave number and sensitive growth rate).

The most unstable wave number versus $Bo$ is demonstrated to verify the accuracy of formulae derived from the asymptotic theory and modal analysis. A good fitting relationship between the modal analysis and the transient growth analysis is illustrated. It is found that an asymptotic behavior exists for both the cylindrical and spherical problems, because the most unstable wave number computed from transient growth analysis are close to the results from modal analysis as $Bo$ increases. Compared to the experimental fingering number for typical coating flows on upper sphere, the agreement between the asymptotic theory and experimental data is good, even though the value of $Bo$ for the experimental cases is not sufficiently large. The linear growth rate is improved observably by the addition of disjoining pressure for cases with the same precursor thickness. But the most unstable wave number is insensitive to the additional disjoining pressure. The same asymptotic behavior in partial wetting cases is observed for both of the cylindrical and spherical problems, which indicates that the asymptotic behavior is not affected by the introduction of wetting model.

The study in current paper can be considered as a preliminary work for understanding the fingering instability on a general two-dimensional or axisymmetric curved surface. Because for high $Bo$ flow, the linear (early) stage of the fingering instability only occurs in a sufficiently small inner region (compared to the whole coating area), thus it is believed that the fingering number may have regularities on a general curved surface which is determined by the local property of a moving front such as front location and thickness. The linear growth on arbitrary curved surface may degenerate into a common well-studied eigenvalue problem if the asymptotic conditions are satisfied, which makes the canonical problem of coating flow on a slope become universal. The fingering number in a low $Bo$ coating flow may deviate observably from the power laws, but the transient growth analysis still holds. It may be a future direction for the research of fingering instability on curved substrates.

\appendix*

\section{\label{adx:num}Numerical methods for eigenvalue problem and linear growth}

The discrete spectra of a linear operator for an unbounded domain are best defined and constructed by the Evans function method. The definition is related to how the eigenfunction behaves at negative and positive infinite far. These asymptotic behaviors are determined by the eigenvalue equation (\ref{eq:eigen}) where the base state $h'_0$ and the coefficients of (\ref{eq:eigen}) asymptotically approach constant values. Since $h'_0$ approaches 1 at $\xi'\rightarrow-\infty$, (\ref{eq:eigen}) degenerates into
\begin{equation}\label{eq:uplimeq}
\frac{\ud^{4}G'}{\ud\xi'^{4}}-2q'^{2}\frac{\ud^{2}G'}{\ud\xi'^{2}}+(2-\delta-\delta^{2})\frac{\ud G'}{\ud\xi'}+(q'^{4}+\beta)G'=0
\end{equation}
The characteristic equation of (\ref{eq:uplimeq}) is
\begin{equation}\label{eq:upchpl}
(\sigma^{-})^4-2q'^{2}(\sigma^{-})^2+(2-\delta-\delta^{2})\sigma^{-}+q'^{4}+\beta=0
\end{equation}
thus the upstream limiting solution $G'\sim e^{\sigma^{-}\xi'}$. Similarly, the degenerate form of (\ref{eq:eigen}) at $\xi'\rightarrow+\infty$ where $h'_0$ approaches $\delta$ is
\begin{equation}\label{eq:dwlimeq}
\delta^{3}\frac{\ud^{4}G'}{\ud\xi'^{4}}-2\delta^{3}q'^{2}\frac{\ud^{2}G'}{\ud\xi'^{2}}+(2\delta^{2}-\delta-1)\frac{\ud G'}{\ud\xi'}+(\delta^{3}q'^{4}+\beta)G'=0
\end{equation}
and the downstream limiting solution $G'\sim e^{\sigma^{+}\xi'}$, in which $\sigma^+$ satisfy the characteristic polynomial
\begin{equation}\label{eq:dwchpl}
\delta^{3}(\sigma^{+})^4-2\delta^{3}q'^{2}(\sigma^{+})^2+(2\delta^{2}-\delta-1)\sigma^{+}+\delta^{3}q'^{4}+\beta=0
\end{equation}

There are four possible complex roots for the quartic equations (\ref{eq:upchpl}) and (\ref{eq:dwchpl}). For a leading unstable (positive) eigenvalue $\beta$, (\ref{eq:upchpl}) yields two roots with positive real parts and two roots with negative real parts (according to Vieta theorem). In contrast, under the same value of $\beta$, (\ref{eq:dwchpl}) yields two roots with positive real parts too. Because all the eigenfunctions must be bounded, only asymptotic decay at $-\infty$ with $\mathrm{Real}(\sigma^-)>0$ and the same behavior at $+\infty$ with $\mathrm{Real}(\sigma^+)<0$ are allowed. Thus there are two possible candidates at both negative (assumed $\sigma^-_1$, $\sigma^-_2$) and positive (assumed $\sigma^+_3$, $\sigma^+_4$) infinities for a given $\beta$. If the upstream limiting solution $e^{\sigma^{-}_1\xi'}$ and $e^{\sigma^{-}_2\xi'}$ are used independently as initial conditions to integrate eigenvalue equation (\ref{eq:eigen}) from $-\infty$, the resulting asymptotic behaviors at $+\infty$ are
\begin{equation*}
G'\sim a_{11}e^{\sigma_{1}^{+}\xi'}+a_{12}e^{\sigma_{2}^{+}\xi'}+a_{13}e^{\sigma_{3}^{+}\xi'}+a_{14}e^{\sigma_{4}^{+}\xi'}
\end{equation*}
and
\begin{equation*}
G'\sim a_{21}e^{\sigma_{1}^{+}\xi'}+a_{22}e^{\sigma_{2}^{+}\xi'}+a_{23}e^{\sigma_{3}^{+}\xi'}+a_{24}e^{\sigma_{4}^{+}\xi'}
\end{equation*}
respectively. To obtain a bounded solution at positive infinity, we need to find a way to linearly combine (because the eigenvalue equation is linear) these two solutions to suppress asymptotic behaviors of $e^{\sigma^{+}_1\xi'}$ and $e^{\sigma^{+}_2\xi'}$, which are unbounded. It is possible if and only if
\begin{equation}\label{eq:evanseq}
E(\beta,q',\delta)=\det(\mathbf{A})=0
\end{equation}
where
\begin{equation}\label{eq:matrixa}
\mathbf{A}=\left(\begin{array}{cc}
                  a_{11} & a_{21} \\
                  a_{12} & a_{22}
                \end{array}\right)
\end{equation}
Thus the eigenvalue $\beta$ is determined by the zeros of the Evans' function $E(\beta,q',\delta)$, these points are then the discrete spectra of the linear perturbation operator with exponential decaying eigenfunctions. In a practical calculation, a guessed value of $\beta$ can be used to calculate the positive roots of (\ref{eq:upchpl}) and hence $e^{\sigma^{-}_1\xi'}$ and $e^{\sigma^{-}_2\xi'}$ at a sufficiently far location upstream, then an iteration is started for eliminating the errors of the Evans' function $E(\beta,q',\delta)$ downstream and computing the accurate value of $\beta$. A numerical shooting method \citep{Press2007} can be used to integrate the eigenvalue equation (\ref{eq:eigen}). Note that the profile of base state $h'_0$ should be integrated as a preliminary to calculate the coefficients in (\ref{eq:eigen}). The adaptive Runge-Kutta solver used to solve the base state can be also applied to obtaining a high-resolution eigenfunction $G'$. Because both the base state and eigenvalue equation are translational invariance in the $\xi'$ direction, the coordinate of downstream point can be translated to a smaller positive value to overcome the stiffness of matrix at positive infinity.

This Evans function shooting method can be extended to (\ref{eq:eigendjp}) which includes the disjoining pressure terms. The main differences compared to (\ref{eq:eigen}) are the characteristic equation at negative and positive infinite far, which are
\begin{multline}\label{eq:upchpldjp}
(\sigma^{-})^4+[K(m\delta^{m}-n\delta^{n})-2q'^{2}](\sigma^{-})^2+(2-\delta-\delta^{2})\sigma^{-}+q'^{4}+\beta-\\
q'^{2}K(m\delta^{m}-n\delta^{n})=0
\end{multline}
at $\xi'\rightarrow-\infty$, and
\begin{multline}\label{eq:dwchpldjp}
\delta^{3}(\sigma^{+})^4+[\delta^{2}K(m-n)-2\delta^{3}q'^{2}](\sigma^{+})^2+(2\delta^{2}-\delta-1)\sigma^{+}+\delta^{3}q'^{4}+\beta-\\
\delta^{2}q'^{2}K(m-n)=0
\end{multline}
at $\xi'\rightarrow+\infty$.

A linear operator is defined to simplify the expression of (\ref{eq:cypertq})
\begin{equation}\label{eq:simpert}
\frac{\partial G}{\partial t}-L_{h_{0}}[G]=0
\end{equation}
The Crank-Nicolson scheme of (\ref{eq:simpert}) can be established on the same adaptive mesh used in two-dimensional capillary wave calculation
\begin{equation}\label{eq:numpert}
\frac{G_{i}^{k+1}-G_{i}^{k}}{\Delta t}-\frac{1}{2}L_{h_{0}\Delta}({h_{0}}_{i}^{k+1},q,\Delta\theta)[G_{i}^{k+1}]-
\frac{1}{2}L_{h_{0}\Delta}({h_{0}}_{i}^{k},q,\Delta\theta)[G_{i}^{k}]=0
\end{equation}
$L_{h_{0}\Delta}$ is the discrete operator using central difference scheme with respect to $L_{h_0}$. The value of ${h_0}_i^k$ can be given using the numerical profile of capillary wave at current time step. For a given wave number $q$, the value of perturbation function $G_i^{k+1}$ at the next time step is calculated by solving a pentadiagonal linear equations system. Finally the numerical profile of linear perturbation at a certain time is obtained. A similar time-marching finite difference method can be derived for (\ref{eq:sppertq}) on spherical surface. There is no complication for the numerical schemes of (\ref{eq:cypertqdjp}) or (\ref{eq:sppertqdjp}) if the disjoining pressure terms are added, because the derivatives of $h_0$ and $G$ do not appear in the disjoining pressure terms and the discrete form can be written directly.

\bibliography{physflupaper}

\end{document}